\documentclass[english]{aa}
\usepackage[T1]{fontenc}
\usepackage[latin1]{inputenc}
\usepackage{graphicx}
\usepackage{amssymb}
\usepackage[authoryear]{natbib}

\makeatletter

\providecommand{\tabularnewline}{\\}

\usepackage{aas_macros}

\usepackage{babel}
\makeatother
\begin{document}

\title{Simulations of the cosmic infrared and submillimeter background for future large
surveys: I. Presentation and first application to Herschel/SPIRE and Planck/HFI}


\author{N. Fernandez-Conde, G. Lagache, J.-L. Puget, H. Dole}

\institute{Institut d'Astrophysique Spatiale (IAS), b\^atiment 121, Universit\'e
Paris-Sud 11 and CNRS (UMR 8617), 91405 Orsay, France e-mail: {[}nestor.fernandez;guilaine.lagache;jean-loup.puget;herve.dole]@ias.u-psud.fr }

\offprints{N. Fernandez-Conde, G. Lagache}

\date{6 June 2007}

\titlerunning{Simulations of the cosmic IR and submm background}

\authorrunning{N. Fernandez-Conde}

\abstract{The coming Planck and Herschel missions will survey the sky at unprecedented
angular scales and sensitivities. Simulations are needed for better
interpretating the results of the surveys and for testing new
methods of, e.g., source extraction and component separation.}
{We present new simulations of the infrared and submillimeter
cosmic background, including the correlation between infrared
galaxies. 
The simulations were used to quantify the source-detection thresholds
for Herschel/SPIRE and Planck/HFI, as well as to study the
detectability of the cosmic infrared background correlated
fluctuations.}
{The simulations are based on an empirical model
of IR galaxy evolution. For these correlations, we only included
the linear clustering, assuming that infrared galaxies
are biased tracers of the dark-matter fluctuation density
field.}
{We used the simulations with different bias parameters to predict
the confusion noise for Herschel/SPIRE and Planck/HFI and the completeness
levels.  We also discuss the detectability of the 
linear clustering in Planck/HFI power spectra, including the
foreground and backgrounds components.}
{Simulated maps and catalogs are publicly available online 
at http://www.ias.u-psud.fr/irgalaxies/simulations.php.
}

\maketitle

\section{INTRODUCTION}

The cosmic infrared background (CIB) ($\lambda\geq8\mu m$) is the
relic emission of the formation and evolution of galaxies. The first
observational evidence of this background was reported by \citet{1996A&A...308L...5P}
and then confirmed by \citet{1998ApJ...508...25H} and \citet{1998ApJ...508..123F}. The 
discovery of a surprisingly high amount of energy in the CIB has shown the importance
of studying its sources to understand how the bulk of stars was formed
in the Universe. Deep cosmological surveys have been carried out thanks
to ISO \citep [see] [for reviews] {2000ARA&A..38..761G,2005SSRv..119...93E}
mainly at 15 $\mu m$ with ISOCAM \citep [e.g.] [] {2002A&A...384..848E};
at 90 and 170 $\mu m$ with ISOPHOT \citep [e.g.] [] {2001A&A...372..364D};
to SPITZER at 24, 70, and 160 $\mu m$ \citep [e.g.] []
{2004ApJS..154...70P,2004ApJS..154...87D} 
and to ground-based instruments such as SCUBA
\citep [e.g.] [] {1998astro.ph..9121H} and MAMBO \citep [e.g.] []{2000astro.ph.10553B}
at 850 and 1300 $\mu m$, respectively. These surveys have allowed
for a better understanding of the CIB and its sources \citep [see]
[for a general review]
{2005ARA&A..43..727L}. 
Some of the results include:
the energy of the CIB is dominated by starbursts although AGN (active
galactic nucleus) contribute too, and the dominant contributors to the
energy output are the LIRGs (luminous IR galaxies) at $z\sim 1$ and
ULIRGs (ultra luminous IR galaxies) at $z\sim 2-3$. \\

Determination of the CIB
by the COBE satellite has been hindered by the accuracy of subtracting
the foreground by only providing just upper limits at 12,
25, and 60 $\mu m$ \citep {1998ApJ...508...25H}, lower limit has
been derived at 24 $\mu m$ by \citet{2004ApJS..154...70P}
as well as the contribution of 24 $\mu m$ galaxies to the background
at 70 and 160 $\mu m$ \citep {2006A&A...451..417D}. The
contribution of the galaxies down to 60 $\mu Jy$ at 24 $\mu m$
is at least 79\% of the 24 $\mu m$ background, and 80\% of the 70 and 160 $\mu m$
background. For longer wavelengths, recent studies have investigated
the contribution of populations selected in the near-IR to the
far-infrared background (FIRB, $\lambda > 200 \mu m$): 
3.6 $\mu m$ selected sources to the 850 $\mu m$ background
\citep {2006ApJ...647...74W} and 8 $\mu m$ and 24 $\mu m$ selected
sources to the 850 $\mu m$ and 450 $\mu m$ backgrounds \citep{2006ApJ...644..769D}.
Similar studies with Planck and Herschel will provide even more evidence
of the nature of the FIRB sources.\\

Studying correlations in the spatial distribution of IR galaxies as
a function of redshift is an essential observation (parallel to
the studies of individual high-redshift, infrared, luminous galaxies), to understand the underlying
scenario and physics of galaxy formation and evolution. A first 
study has been done using the 850 $\mu$m galaxies
\citep{2004ApJ...611..725B}. Although the number of sources is quite
small, they find evidence that submillimiter galaxies are
linked to the formation of massive galaxies in dense environments
destined to become rich clusters. This has now been directly supported
by the detection of the clustering of high-redshift 24 $\mu$m selected
ULIRGs and HyperLIRGs \citep{2006ApJ...641L..17F,2007MNRAS.375.1121M}. Studying
correlations with individual IR galaxies is very hard due to either
high confusion noises, instrumental noises, or small fields of observation. 
It has been shown that the IR-background anisotropies could provide
information on the correlation between the sources of the CIB and
dark matter for large-scale structures \citep [] [hereafter HK] {2001ApJ...550....7K,2000ApJ...530..124H} 
and on the large-scale structure
evolution. First studies at long wavelengths 
have only detected the shot-noise component of the fluctuations: 
\citet{2000A&A...355...17L} at 170 $\mu m$, \citet{2000A&A...361..407M}
at 90 and 170 $\mu m$, \citet{2001phso.conf..471M} at 60 and 100
$\mu m$.  \citet{2007ApJ...665L..89L} and \citet{2007A&A..4512G}
report first detections of the correlated component using
Spitzer/MIPS data at 160 $\mu m$. \citet{2007ApJ...665L..89L} measured a linear bias $b \sim 1.7$.
\\

Future observations by Herschel and Planck will allow us to probe
the clustering of IR and submm galaxies. 
Nevertheless these experiments will be limited, the confusion
and instrumental noises will hinder detections of faint individual
galaxies.
Clustering thus has to be analysed in the background fluctuations
\citep [e.g.] [] {2007astro.ph..3210N}.
The need for a prior understanding of what could be done by these experiments
has motivated us to develop a set of realistic simulations of the IR and sub-mm sky. \\

In Sect. 2 we present the model on which are based our simulations.
In Sect. 3 we discuss how the simulations are done
and present a set of simulated sky maps and their corresponding
catalogs. Different catalogs are created for 3 different levels of
correlation between the IR galaxy emissivity
and the dark-matter fluctuation density field (strong, medium, and no correlation). For each of these catalogs, we can create maps
of the sky at any given IR wavelength and simulate how different instruments
will see them. We focus in this paper on Planck/HFI and Herschel/SPIRE.
In Sect. 4 we use the simulated maps
to give predictions for the confusion noise, the completeness, and
the detection limits for each of the study cases, including  the instrumental
noise. In Sect. 5 we present the power spectra of the
CIB anisotropies 
for Planck/HFI and discuss
their detectability against the significant sources of contamination
(shot noise, cirrus, and cosmic microwave background
(CMB)). \\

Throughout the paper the cosmological parameters were set
to $h=0.71,\Omega_{\Lambda}=0.7,\Omega_{m}=0.27$. For the dark-matter
linear clustering we set the normalization to $\sigma_{8}=0.8$.

\section{\label{model} THE MODEL}

\subsection{Galaxies' empirical evolution model}

The model of IR galaxies used for the simulations is from \citet{2003MNRAS.338..555L},
revisited in \citet{2004ApJS..154..112L} -- hereafter the LDP model,
see http://www.ias.u-psud.fr/irgalaxies/model.php.
This model is a flexible tool for planning surveys and developing analysis
methods. The requirement was to build the simplest model of the luminosity
function (LF) evolution with redshift, with the lowest number of parameters,
but accounting for all statistical observational data between 5 $\mu m$
and 1 mm. These are the spectral
energy distribution of the CIB and its fluctuations, galaxy luminosity
functions and their redshift evolution, as
well as the existing source counts and redshift distributions.\\

The luminosity function of IR galaxies was modelled by a bimodal
star-formation process: one associated with the passive phase of 
galaxy evolution (normal galaxies) and one associated with the starburst
phase, mostly triggered by merging and interactions (starburst galaxies).
Unlike for the starburst galaxies, the normal galaxy contribution to
the luminosity function was considered
mostly unchanged with redshift. The
spectral energy distribution (SED) changes with the luminosity of
the source but is assumed constant with redshift for both populations
in this simple model.
\\

\begin{figure}
\includegraphics[width=1\columnwidth]{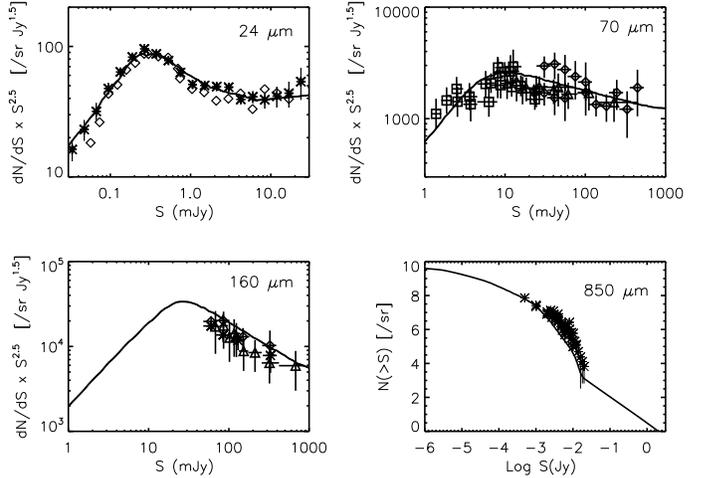}

\caption{Comparison of the observed source counts (data points) and model
predictions (continuous lines). {\it Upper left} 24 $\mu$m,  {\it
Upper right} 70 $\mu$m,  {\it Lower left} 160 $\mu$m, {\it Lower right} 850 $\mu$m.
\label{fig: Source Counts}}
\end{figure}

This model fits all the experimental data and has predicted that LIRGs
($10^{11}<L_{IR}<10^{12}$) dominate at $z\simeq0.5-1.5$ and that ULIRGs/HLIRGs ($L_{IR}>10^{12}$) 
dominate at $z\simeq2-3$ 
the energy distribution of the CIB.
One example of the agreement between the model
and the observations is shown in Fig. \ref{fig: Source Counts}.

\subsubsection{Number counts and CIB fluctuations}

To illustrate the interest of studying the cosmic background
fluctuations in the far-IR and submm domains, we use
a simplistic approach for the number counts, following \citet{2000A&A...355...17L}.
The source number counts can be schematically represented
by a power law:

\begin{equation}
N(S>S_{0})=N_{0}\left(\frac{S}{S_{0}}\right)^{-\alpha}\label{eq:N(S)}
\end{equation}

where we set $S_{0}$ to be the detection limit for the sources and $N_{0}$
the number of sources with flux larger than $S_{0}$.\\

In a Euclidean Universe with uniform density of the sources
$\alpha=1.5$. In the far-IR and submm, a steeper slope is observed
with $\alpha=2-3$ in the regime where negative K-correction dominates.
As an example, ISO observations found a slope of $\alpha=2.2$ at 170
$\mu m$ \citep {2001A&A...372..364D}. Obviously, the number counts need to
flatten for low fluxes to ensure that the CIB remains finite. For the
rest of the discussion we will assume that $\alpha=0$ for $S<S^{*}$. 
The total intensity of the CIB composed by all the sources up to $S_{max}$
is given by:

\[
I_{CIB}=\int_{0}^{S_{max}}{\displaystyle S\frac{dN}{dS}dS}.\]

For the Euclidean case the CIB intensity is dominated by sources near
$S^{*}$.\\

Fluctuations from sources below the detection limit $S_{0}$ are given
by

\[
\sigma^{2}=\int_{0}^{S_{0}}S^{2}\frac{dN}{dS}dS.\]

Using $\frac{dN}{dS}$ given by Eq. \ref{eq:N(S)} we get\\

\[
\sigma^{2}=\frac{\alpha}{2-\alpha}N_{0}S_{0}^{2}\left[1-\left(\frac{S^{*}}{S_{0}}\right)^{2-\alpha}\right].\]

\begin{figure}
\includegraphics[width=1\columnwidth]{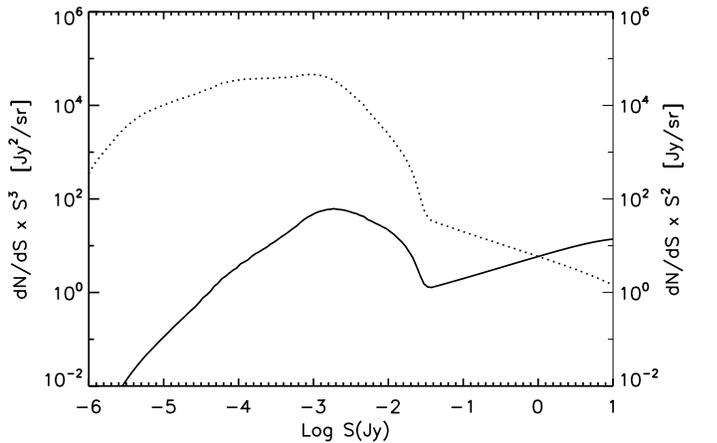}

\caption{Contributions of the sources of flux S (in Jy) per Log interval of S to the background
(dotted line, right y axis) and fluctuations (continuous line, left y
axis) at $850\mu m$. \label{fig: Fluctuations-BackGround-S}}
\end{figure}

For $\alpha>2$ CIB fluctuations are dominated by sources close to $S^{*}$
so that the same sources dominate both the FIRB and its fluctuations. Therefore
by studying the fluctuations of the FIRB, we are also studying the
sources that form the bulk of the contribution to the FIRB. We can
check this conclusion with the number counts from the LDP model.
Figure \ref{fig: Fluctuations-BackGround-S} shows that the same sources
dominate the background and the fluctuations, but only for faint sources
(for example $S_{850}\lesssim50$ mJy). Therefore, it is necessary 
to subtract bright sources prior to any fluctuation analysis 
since they would otherwise dominate the fluctuations.

\subsubsection{IR galaxy emissivity}

For the purpose of the model we need to compute the mean IR galaxy
emissivity per unit of comoving volume $[W/Mpc^{3}/Hz/sr]$. It is
defined as

\begin{figure}
\includegraphics[clip,width=1\columnwidth]{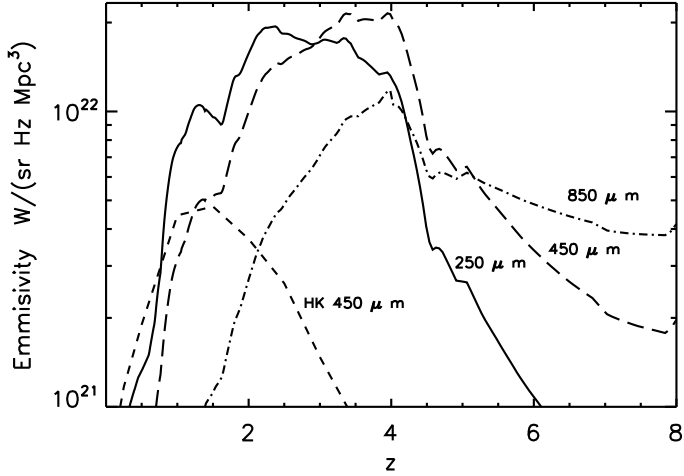}

\caption{Emissivities computed using the LDP model at (observed) 250 $\mu m$ (continuous line),
450 $\mu m$ (long-dashed line), and 850 $\mu m$ (dotted-dashed
line). The emissivity from HK at 
450 $\mu m$ (short-dashed line) is shown for comparison .\label{cap:Emissivities}}
\end{figure}

$j_{d}(\nu, z)=(1+z)\int_{L_{bol}}L_{\nu'=\nu(1+z)}\frac{dN}{dln(L_{bol})}dln(L_{bol})$

where L is the luminosity (in $W/Hz/sr$), $\frac{dN}{dln(L_{bol})}$is
the comoving luminosity function (in $Mpc^{-3}$), and $\nu$ the
observed frequency. We compute $j_{d}$ using the SEDs and luminosity
function from the LDP model which assumes that the SED depends only on
$L_{bol}$. 
The resulting $j_{d}$ is different from
what is used by former approaches \citep [HK,] [] {2001ApJ...550....7K}.
We can see  the difference between
the emissivity from our model and that of HK in Fig. \ref{cap:Emissivities}. The crude
model used for the emissivities by HK gives 
much lower emissivities than ours.\\

\subsection{IR galaxy spatial distribution}

Any model trying to account for CIB fluctuations must describe the statistical properties of 
the spatial distribution of the sources. The absence of a completely
developed theoretical model for the distribution of the whole IR
galaxy populations
makes the empirical modelling of different distributions for the sources
of the CIB necessary in order to prepare future observations.
We used 
an empirical description for the spatial distribution of these sources,
which has been used to create the simulated sky maps. \\

The LDP model did not address the spatial distribution problem due
to the lack of constraints at the time it was built. This is still
mostly the case at the time of writing this work. The simulations by \citet{2003ApJ...585..617D}
did not implement any correlation between IR galaxies and used 
an uncorrelated random distribution. However, since future experiments
such as Herschel and Planck will be able to detect large-scale
IR galaxy correlations ($\ell \lesssim 1000$), 
a model addressing this problem has
become necessary. Herschel, with its high angular resolution, is expected to also probe
correlations between galaxies in the same dark-matter haloes but in
this study this correlation has not been considered for simplicity.
We only consider the linear clustering i.e. IR galaxies
as biased tracers of the dark matter haloes, with a linear relation
between the dark-matter density-field fluctuations and IR emissivity.\\

We follow the prescription from \citet{2001ApJ...550....7K}. The
angular power spectrum that characterises the spatial
distribution of the fluctuations of the CIB can be written as

\begin{equation}
C_{l}^{\nu}=\int\frac{dz}{r^{2}}\frac{dr}{dz}a^{2}(z)\bar{j}^2_d(\nu,z)b^2(k,\nu,z)P_{M}(k)\vert_{k=l/r}G^{2}(z).
\label{eq:Angular power spectrum}
\end{equation}

In the equation several components can be identified, starting with 
a geometrical one $\frac{dz}{r^{2}}\frac{dr}{dz}a^{2}(z)$ (these
terms take  all the geometrical effects into account), followed by 
the galaxies emissivity $\bar{j}_d(\nu,z)$ already described in Sect. 2.1.2, 
then the bias $b(k,\nu,z)$, and finally the power
spectrum of dark-matter density fluctuations today
$P_{M}(k)\vert_{k=l/r}$
and the linear theory growth function $G^{2}(z)$. Finally
$\ell$ is the angular multipole, in the Limber approximation $k=l/r$, and r the proper motion
distance. 
The way the power spectrum has been obtained is developed
in the following subsections.

\begin{figure}
\includegraphics[width=1\columnwidth]{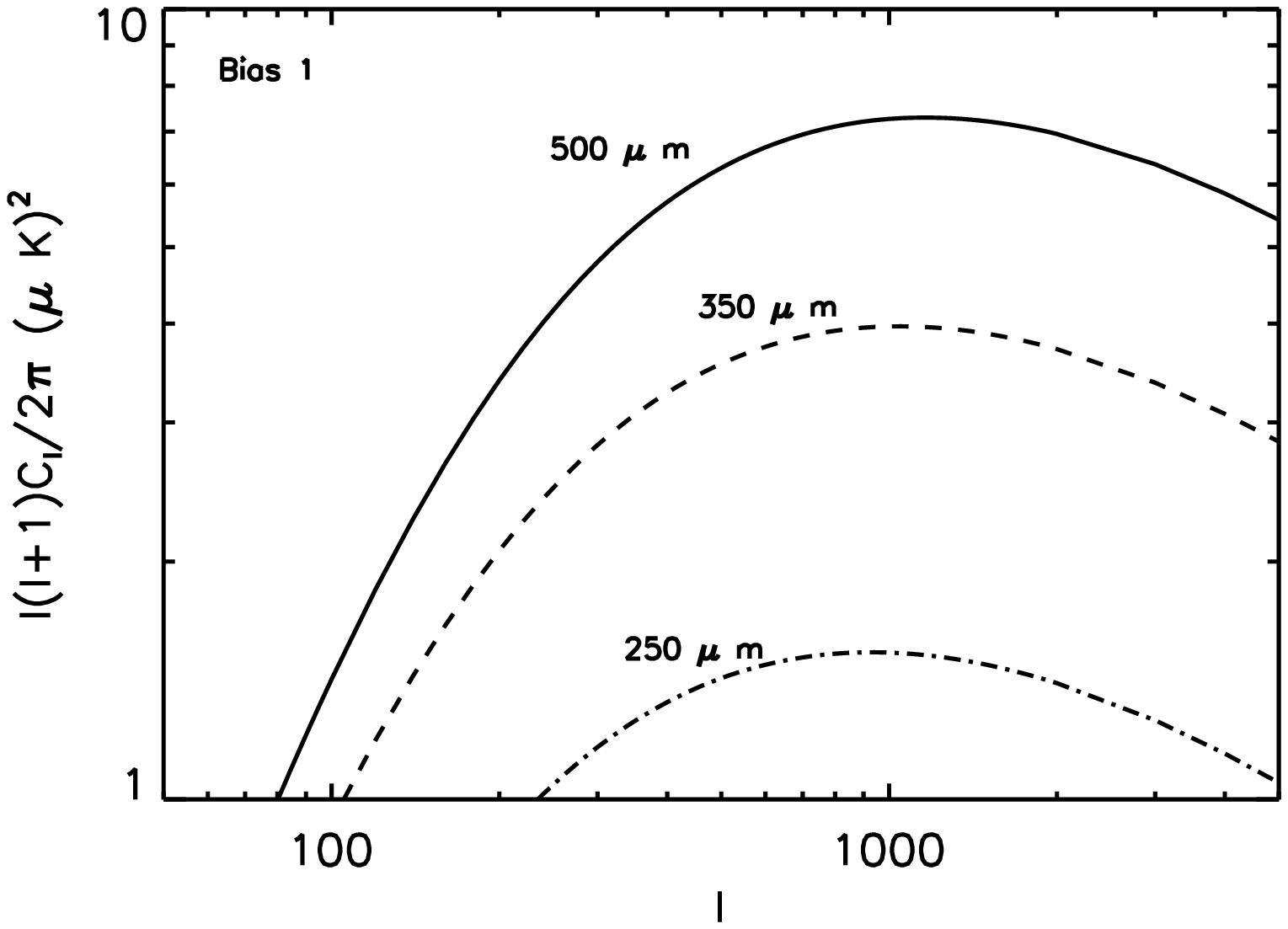}

\includegraphics[width=1\columnwidth]{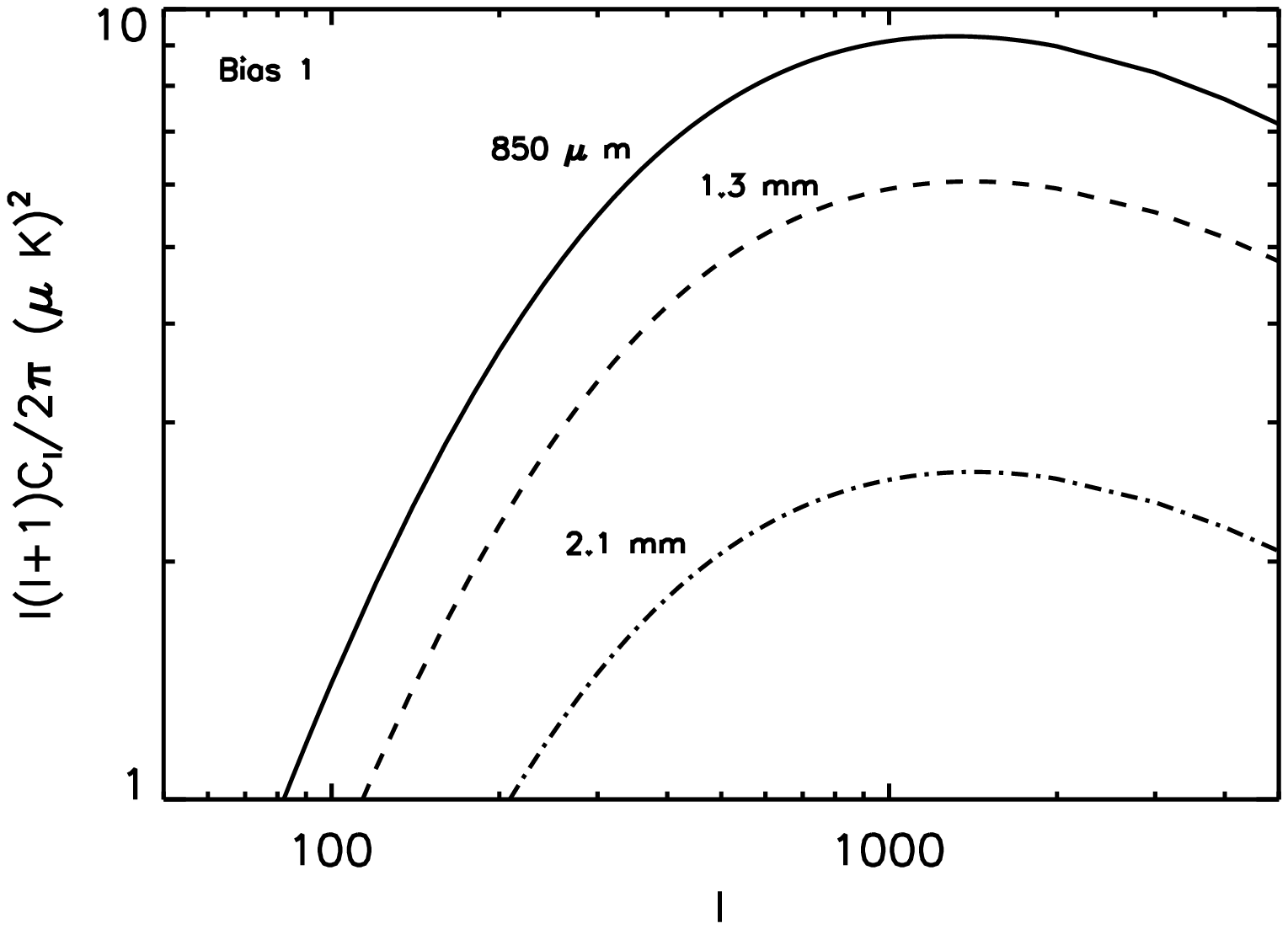}

\caption{Top: CIB Power spectrum with a bias $b=1$ at Herschel/SPIRE wavelengths 500 $\mu m$ (continuous
line), 350 $\mu m$ (dashed line), and 250 $\mu m$(dotted-dashed line)\label{fig:Pk Spitzer-Planck}.}

Bottom: CIB Power spectrum with a bias $b=1$ at Planck/HFI CIB wavelengths 850 $\mu m$ (continuous
line),1380 $\mu m$ (dashed line), and 2097 $\mu m$(dotted-dashed
line). 
\end{figure}

\subsubsection{Dark-matter power spectrum\protect \\
}

The power spectrum of the dark-matter distribution at $z=0$ can be
written as

\begin{equation}
P_{M}(k)\propto kT^{2}(k)
\label{eq:Power spectrum dark matter and transfer function}
\end{equation}

where $T(k,t)$ is the transfer function for a cold dark-matter
universe \citep {1986ApJ...304...15B}.
The linear theory growth function $G(z)$ writes as

\begin{equation}
G^{2}(z)=\frac{g^{2}(\Omega(z),\Omega_{\Lambda}(z))}{{g^{2}(\Omega_{0},\Omega_{\Lambda0})(1+z)^{2}}}\label{eq:Growth factor}\end{equation}

with

\begin{eqnarray*}
g[\Omega(z),\Omega_{\Lambda}(z)] & = & \frac{5}{2}\Omega(z)\times\left[\Omega(z)^{4/7}-\Omega_{\Lambda}(z)\right.\\
 &  & \left.+\left({1+\frac{1}{2}\Omega(z)}\right)\left({1+\frac{1}{70}\Omega_{\Lambda}(z)}\right)\right]^{-1}.\end{eqnarray*}

And $\Omega_{\Lambda}(z)={\frac{1-\Omega_{0}}{\Omega_{0}(1+z)^{3}+1-\Omega_{0}}}$,
$\Omega(z)=\frac{\Omega_{0}(1+z)^{3}}{\Omega_{0}(1+z)^{3}+1-\Omega_{0}}$\\

\subsubsection{Bias Model\protect \\
}

The bias of IR galaxies represents their level of correlation
with the dark-matter density field. It can be expressed as a function of the spatial
scale, the redshift, and the wavelength of observation. In this paper
and due to lack of measurements for the bias, we
consider a simplified constant bias $b$.

\[
\frac{\delta j_{d}(k,\nu,z)}{\bar{j_{d}}(k,\nu,z)}=b\frac{\delta\rho(k,\nu,z)}{\bar{\rho}(k,\nu,z)}\]

where $j_{d}$ is the emissivity of the IR galaxies per comoving unit
volume, $\bar{j_{d}}$ its mean level, and $\delta j_{d}$ its fluctuations.
Similarly, $\rho$ is the dark matter density, $\bar{\rho}$ its
mean value, and $\delta\rho$ is the linear-theory dark-matter density-field
fluctuation. \\

We have better knowledge of the bias for optical and radio galaxies
than for IR galaxies. Several studies have been able to measure
the bias for the optical sources. As an example, a high bias ($b\sim3$)
has been found at $z\sim3$ for the Lyman-Break Galaxies 
\citep{1998astro.ph.12167S,1998ApJ...503..543G,1998ApJ...505...18A}). It has
been found as well that the bias increases with redshift both for
the optical \citep{2006astro.ph.12123M} and the radio \citep{2003NewAR..47..325B}
populations. The optical or radio bias could be misleading as a first
guess for the bias of IR galaxies. IRAS has measured a low bias
of IR galaxies at $z\sim0$ \citep [e.g.] [] {1992MNRAS.258..134S}.
Such a low bias is expected since the
starburst activity in the massive dark-matter haloes in the local
universe is very small. But we expect a higher IR bias at higher $z$, during the epoch
of formation of galaxy clusters. Indeed, \citet{2007ApJ...665L..89L} report 
the first measurements of the bias, $b \sim 1.7$ in the CIB
fluctuations at 160 $\mu m$ using Spitzer data. The LDP model
indicates that galaxies dominating the 160 $\mu$m anisotropies
are at $z\sim 1$. This implies that infrared galaxies at high
redshifts are biased tracers of mass, unlike in the local Universe.
For an extensive review of the bias
problem see \citet{2004LRR.....7....8L}.\\

The IR bias could have very complex functional dependences, namely
with the spatial frequency $k$, the redshift z, and the radiation
frequency $\nu$ (for example if different populations of galaxies
with different SEDs have different spatial distributions). 
However, for the simulations, simplified guesses for the bias were used, namely a constant
bias of 1.5, 0.75, and 0. \\

Figures \ref{fig:Pk Spitzer-Planck} show the angular power spectrum
$C_{l}$ for some Herschel and Planck wavelengths. The power spectra
are shown for a constant bias $b=1$. Since $C_{\ell}\propto b^2$, the predicted
power spectrum scales as $b^2$.

\subsection{Discussion and implications of the model}

\begin{figure}
\includegraphics[width=1\columnwidth]{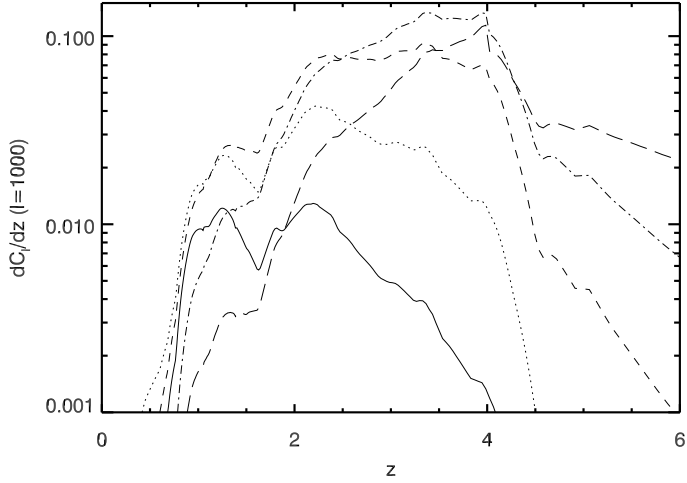}

\caption{Redshift contributions to the angular power spectrum
$\frac{dC_{l}}{dz}$ at $\ell=1000$ in $\mu K^2$ for different wavelengths:
250 $\mu m$ (continuous
line), 350 $\mu m$ (dotted line), 550 $\mu m$ (dashed line), 850
$\mu m$ (dotted-dashed line), and 1380 $\mu m$ (long dashed line).
\label{fig:dCl/dz}}
\end{figure}

\begin{figure}
\includegraphics[width=1\columnwidth]{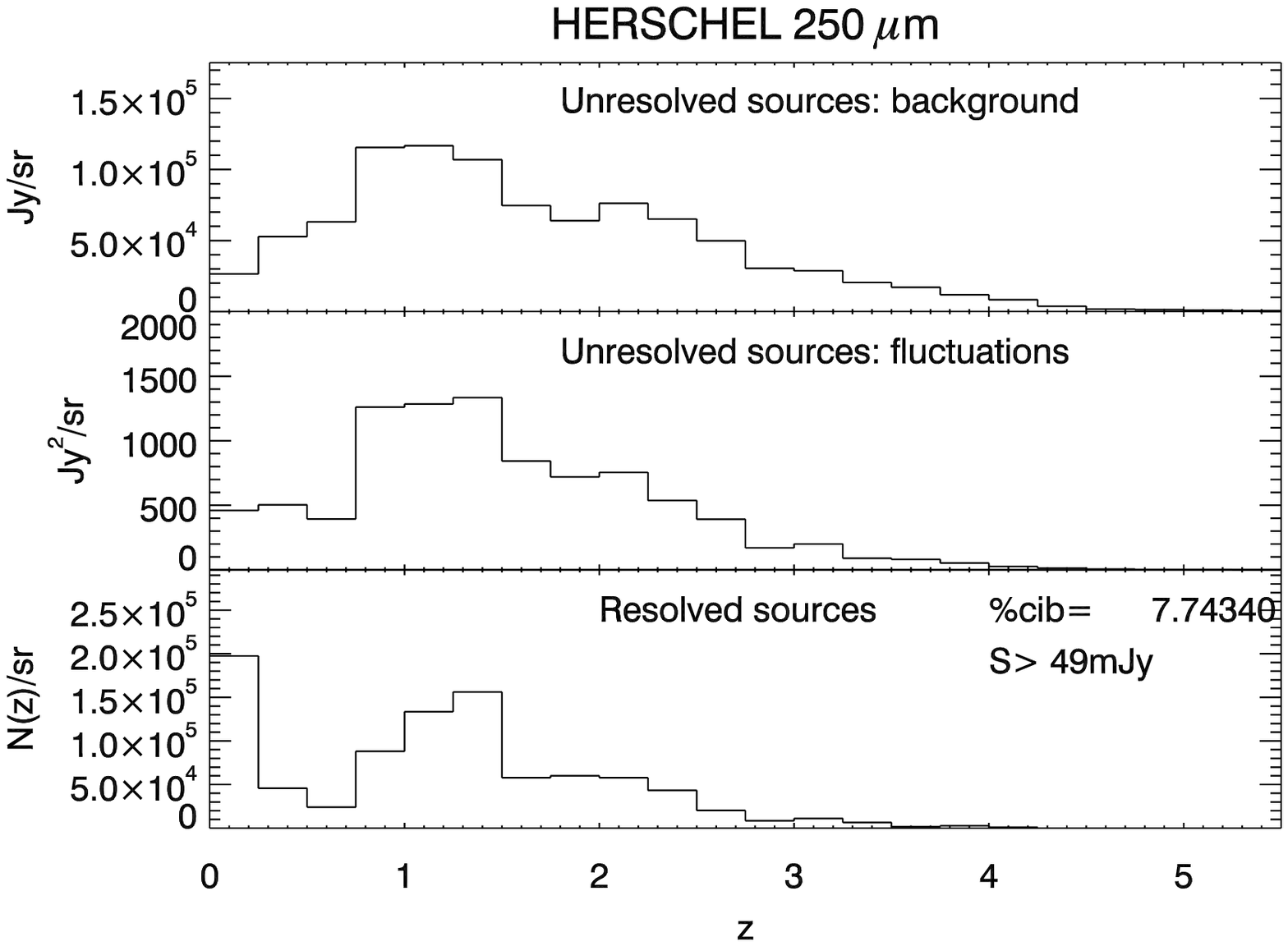}

\includegraphics[width=1\columnwidth]{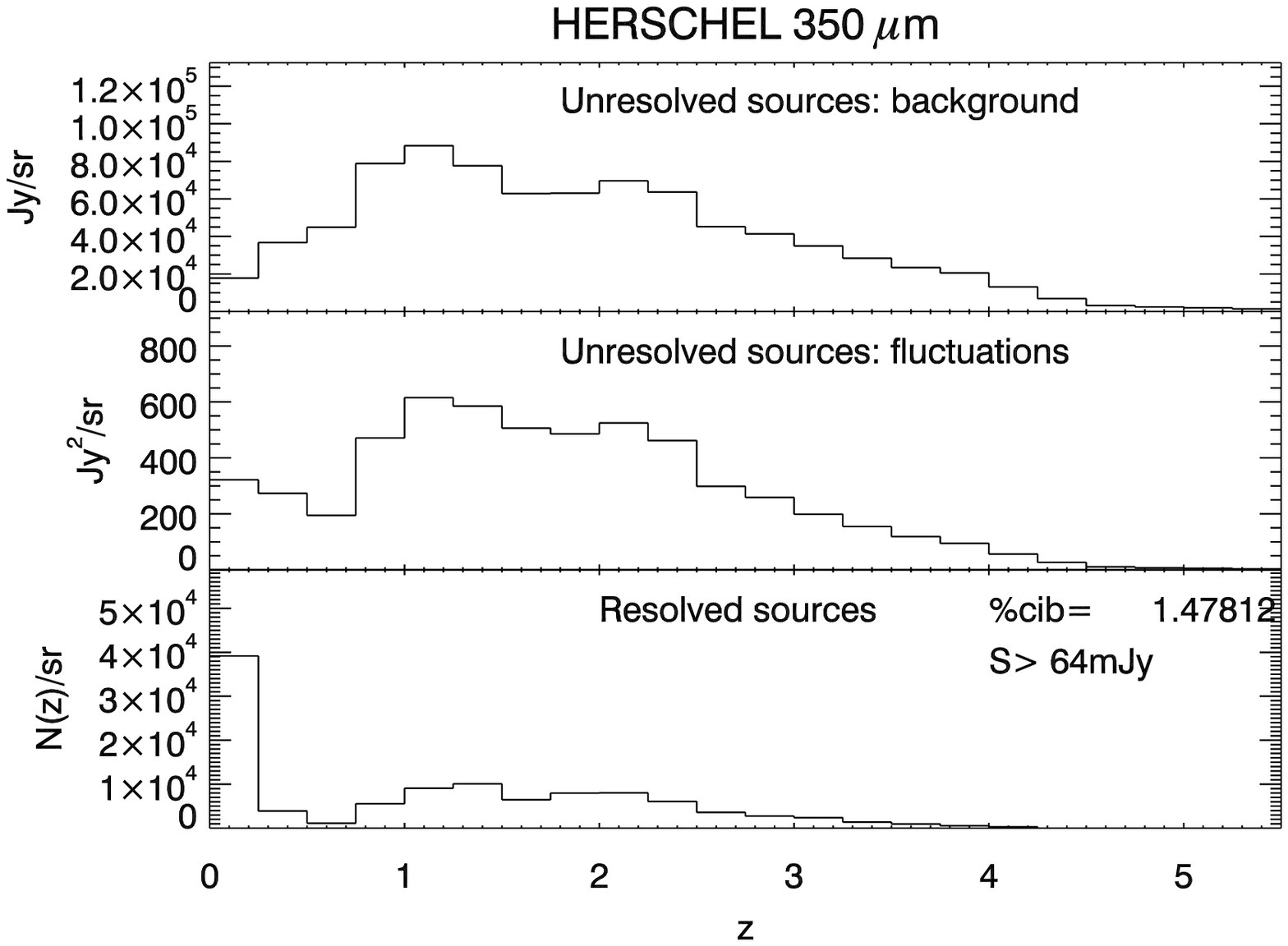}

\includegraphics[width=1\columnwidth]{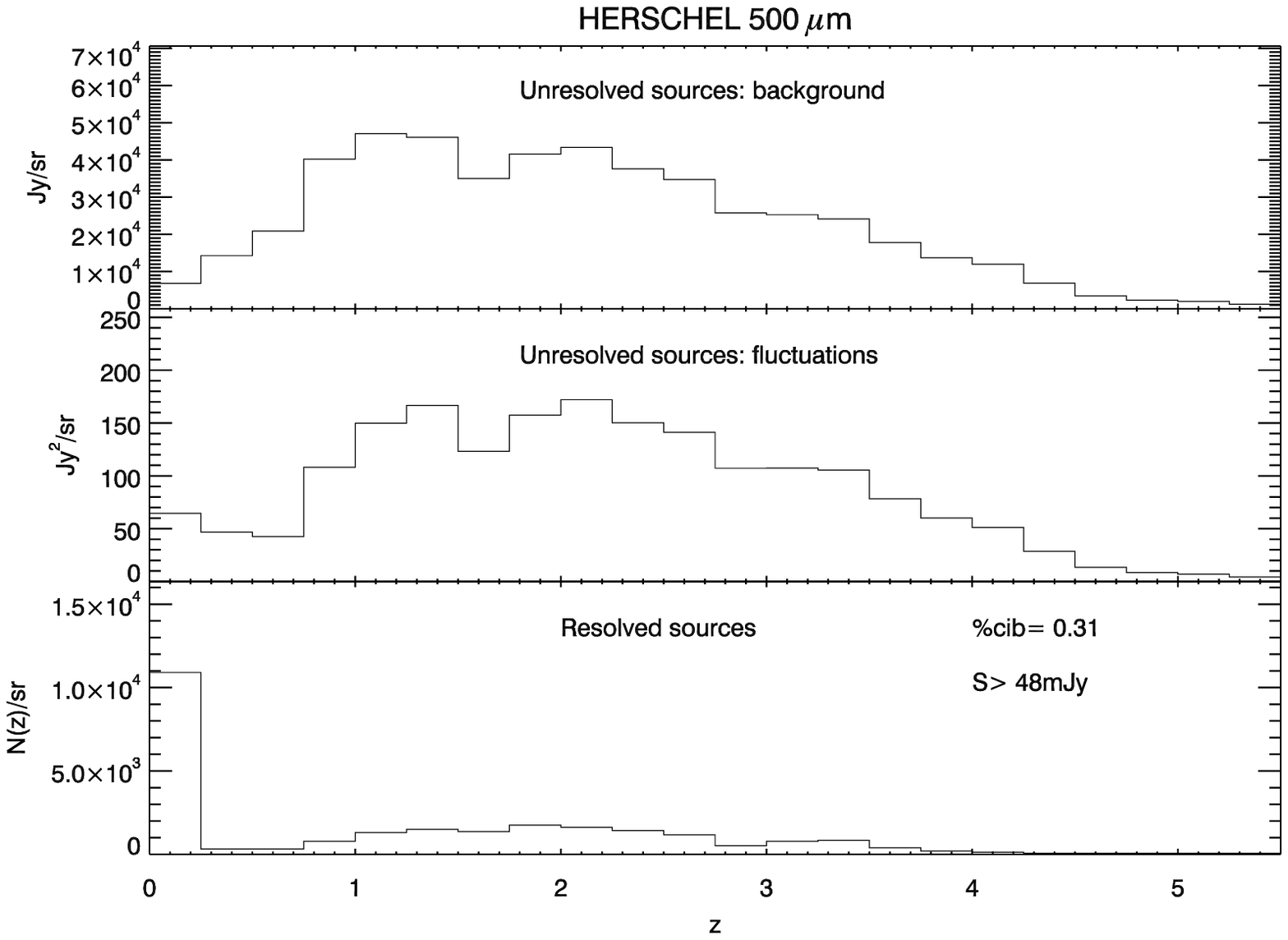}

\caption{Redshift contribution to the FIRB ({\it top panels}) and its
fluctuations ({\it middle panels}).
Also shown are the redshift distributions
of the detected sources ({\it bottom panels}) for a typical
large Herschel/SPIRE deep survey (see Sect 4.3).
From top to bottom: $250 \mu m$, $350\mu m$ and $500 \mu m$.
\label{fig:Z-Fluct-CIB-Sources-Herschel}}
\end{figure}

\begin{figure*}

\begin{centering}\includegraphics[width=1\columnwidth,keepaspectratio]{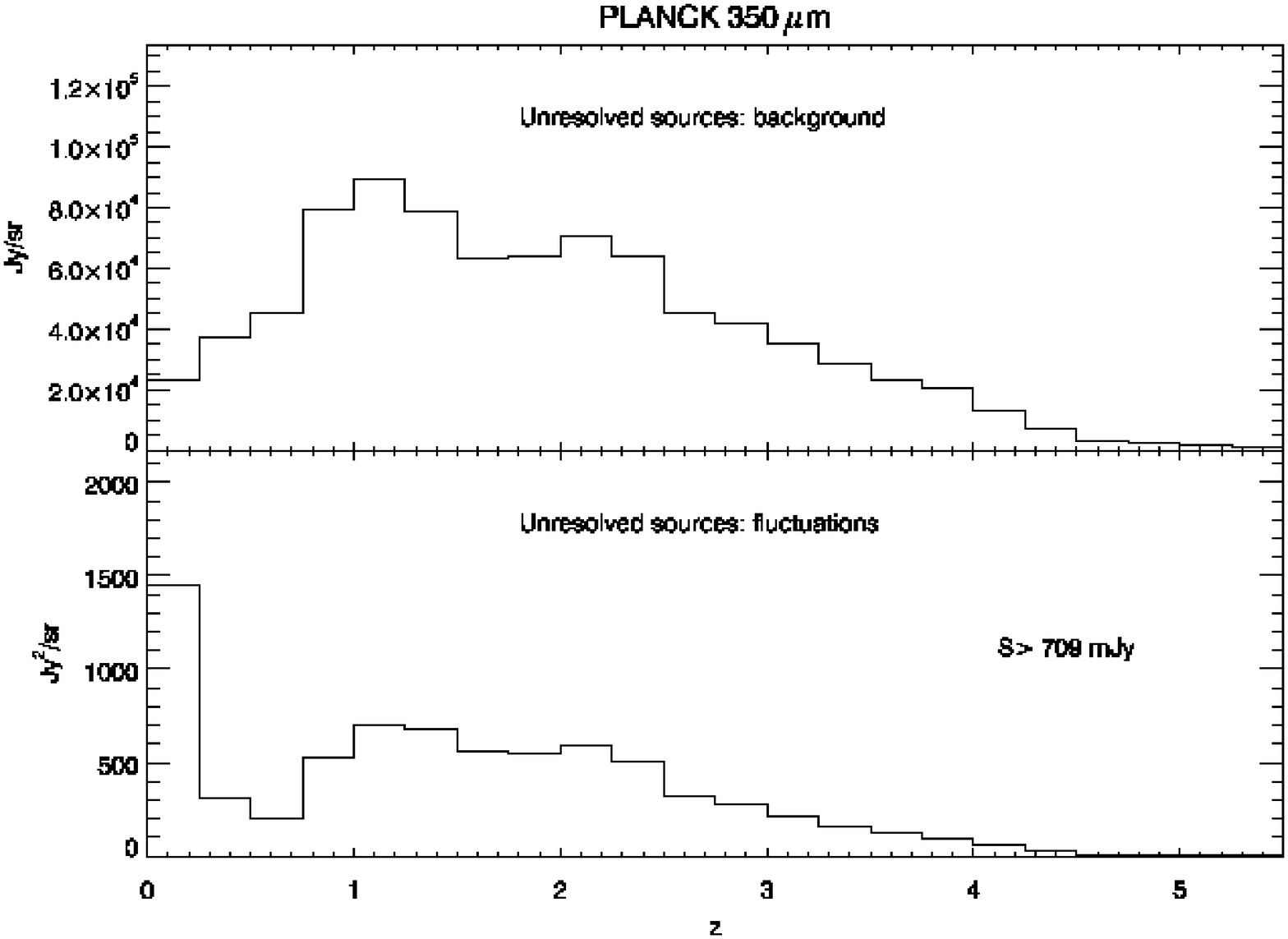}\includegraphics[width=1\columnwidth,keepaspectratio]{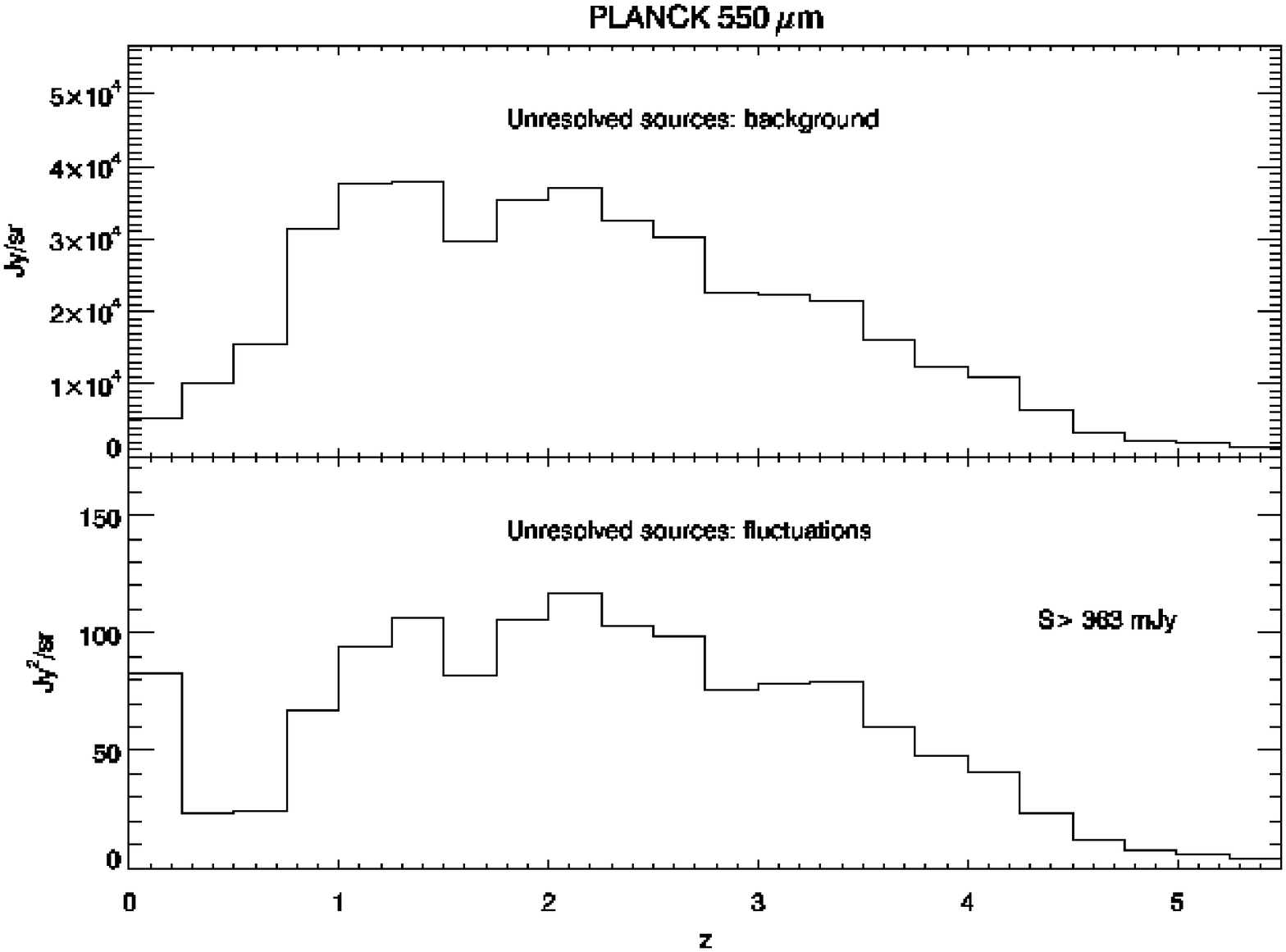}\par\end{centering}

\begin{centering}\includegraphics[width=1\columnwidth,keepaspectratio]{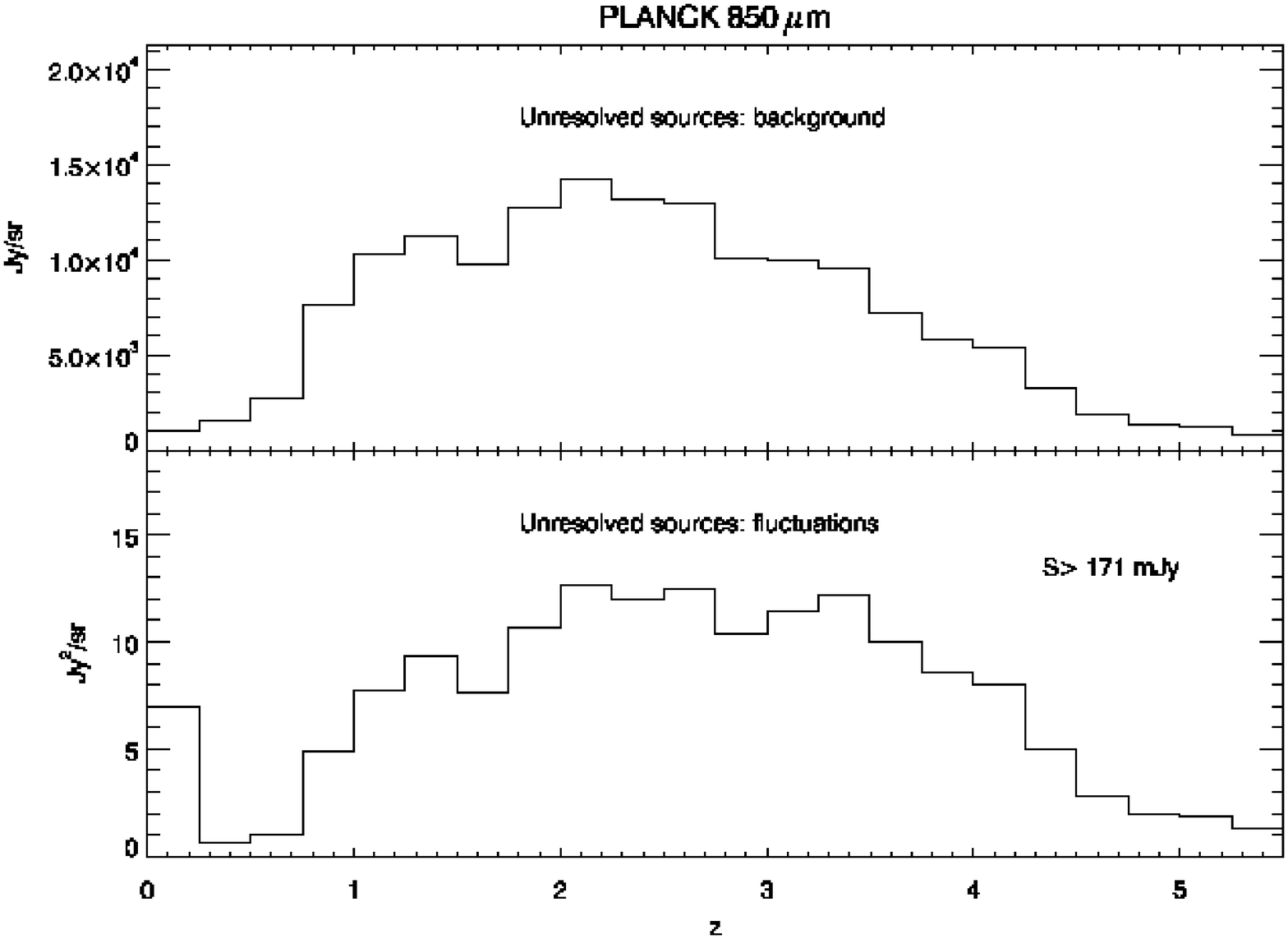}\includegraphics[width=1\columnwidth,keepaspectratio]{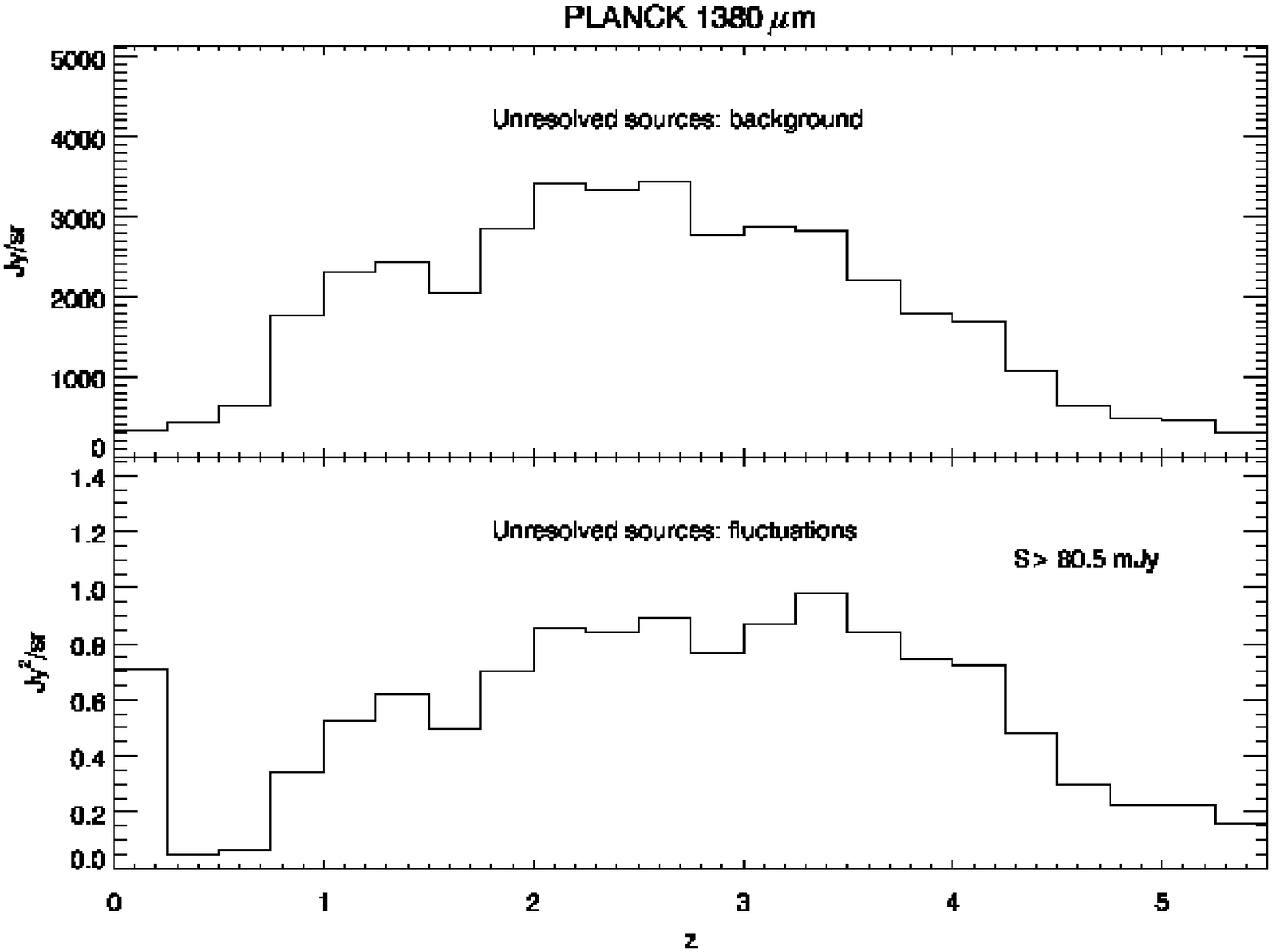}\par\end{centering}

\caption{Redshift contribution to the FIRB ({\it top panels}) and its
fluctuations ({\it bottom panels}) for a Planck simulation (dz=0.25) at
$350 \mu m$ (top-left figure), $550 \mu m$ (top-right figure),
$850 \mu m$ (bottom-left figure), $1380 \mu m$ (bottom-right figure). 
The plots are for simulations with $b=1.5$, which sets
the detection limit (see Sect 4.3). \label{fig:Z-Fluctuations-CIB-Sources-Planck}}
\end{figure*}

The IR galaxy SED peaks near 80 $\mu m$. This combines with the
Doppler shift and causes observations at different wavelengths to probe
different redshifts. Figure \ref{fig:dCl/dz} shows the contributions
to the power spectrum at $l=1000$ for different redshifts, normalized to unity. The contributions
to the same $\ell$ come from higher redshift as wavelengths increase.
The shorter wavelengths probe the lower redshifts because they are
close to the maximum of the SED, while the longer wavelengths probe
the higher redshifts due to the strong negative K-correction. \\

Figures \ref{fig:Z-Fluctuations-CIB-Sources-Planck} and \ref{fig:Z-Fluct-CIB-Sources-Herschel}
show the redshift contributions to the intensity of the CIB and to its integrated rms fluctuations for
Planck/HFI and Herschel/SPIRE, assuming sources with $S>S_{det}$ have been
removed -- $S_{det}$ corresponds to the source detection thresholds
computed in Sect. 4.3. We see that the fluctuations and the FIRB are
dominated by sources at the same redshift. Therefore, studying the
fluctuations at different wavelengths will allow us to study the spatial
distribution of the sources forming the FIRB at different redshifts.\\

The amount of fluctuations that come from sources at redshifts
lower than 0.25 for the Planck/HFI case at 350 $\mu m$ is
noticeable from  Fig. \ref{fig:Z-Fluctuations-CIB-Sources-Planck}.
This contrasts with the
Herschel/SPIRE predictions where the bulk of the low-z sources
contributing to the fluctuations in the Planck case are resolved. These 
individual detections with Herschel/SPIRE could allow their subtraction in the
Planck maps. A similar approach could be used between the Herschel
500 $\mu m$ and the Planck 550 $\mu m$ channels, although it is more
marginal. Using information on the fluctuations at shorter wavelengths
to remove the low-z fluctuations from longer wavelength maps
could be another approach to studying the fluctuations
at high redshifts directly. \\

\begin{figure}
\includegraphics[width=1\columnwidth]{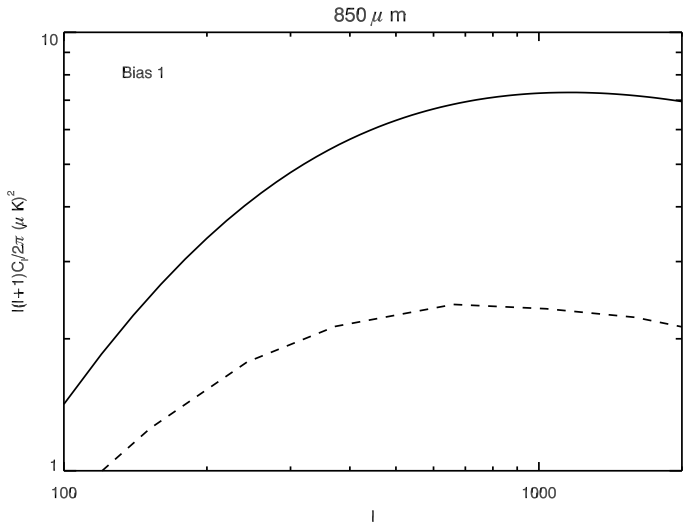}

\caption{Our power spectrum at 850 $\mu m$ (continuous line) and that of \citet{2000ApJ...530..124H}
(dashed line). The differences between both models arise from the
differences in the emissivities (see
Fig. \ref{cap:Emissivities}).
 The lower level of the HK power spectrum comes from their lower
emissivities. 
The emissivities of HK are at lower z and therefore favour larger angular scales for the power spectrum relative to our model.\label{fig:OurCl-vs-HK}}
\end{figure}

A similar model has been developed by HK and revisited by \citet{2001ApJ...550....7K}.
We compare  the HK and our $C_{l}$
prediction at 850 $\mu m$ in Fig. \ref{fig:OurCl-vs-HK} (for the
comparison, the same bias and  $\sigma_{8}$ is used). Our model is 2
times higher mainly due to our
higher prediction for the IR galaxy emissivity. Similar results
are found for other wavelengths.

\section{\label{simulations} THE SIMULATIONS}

The simulations were computed by an IDL program that calculates the
dark-matter power spectrum and spreads the galaxies in the map according
to their correlation with the dark-matter density field. The CIB power spectrum
is calculated as explained in Sect. 2.\\

To create the maps, two assumptions were made: first that all the
galaxies share the same spatial distribution independently of their
luminosities; 
second that both IR and normal galaxies share the same spatial
distribution. 
This second assumption was made to avoid too many free parameters
in the simulations, the contributions of both populations being well
separated in redshift this assumption is a weak one.\\

The process for the creation of a virtual catalog can be summarised
as follows. For a given wavelength, we create the map as a superposition
of maps at different redshifts from $z=0$ to $z=6$. The separation
in redshift slices decorrelates the emission from very
distant regions of the modelled volume of the universe. In order to do so, we divided the maps in
slices covering $dz=0.1$. We can see the
size of these slices for different redshifts in
Table \ref{tab:Z-Slices}. For
all redshift ranges the size of the slices is bigger than the measured comoving correlation
lengths (for all populations of galaxies). We then construct a
brightness map for each redshift slice by adding: 1) a constant map
with the mean surface brightness predicted by the LDP model for that
$z$ slice, 2) a map of the fluctuations for the given bias predicted
by our spatial distribution model for that $z$ slice. The fluctuations
are not correlated between $z$ slices. The brightness map is then converted
into flux map. At each luminosity, this can be converted into maps of
numbers of sources. These numbers of sources are then redistributed into 
smaller $z$ slices (inside the 0.1 slice) to refine the luminosity/flux
relation. Note that all sources have the same underlying low frequency spatial distribution
(but not the same positions) per $dz=0.1$ slice. The position, luminosity,
type (normal or starburst) and redshift of all sources are stored
in a catalog. Since we know these four parameters for all the sources, 
we can now create maps of the sky at any given wavelength. To simulate
the observations, the map is convolved with the point spread function
(PSF) of the chosen instrument.
\\

\begin{table}

\caption{Physical size of the redshift slice $dz=0.1$ (in Mpc) for different z\label{tab:Z-Slices}.}

\begin{tabular}{ccccc}
\hline 
z&
1.0-1.1&
2.0-2.1&
3.0-3.1&
4.0-4.1\tabularnewline
\hline 
$R_{dz=0.1}$ (Mpc)&
233&
139&
93&
67\tabularnewline
\hline
\end{tabular}

\end{table}

\begin{table}

\caption{FWHM of the PSF for different wavelengths of observation (in
arc seconds) for all the simulated maps. \label{cap:Wavelengths and psf of maps}}

\begin{tabular}{cccccc}
\hline 
Wavelengths $(\mu m)$&
350&
550&
850&
1380&
2097\tabularnewline
\hline 
Planck HFI FWHM ('')&
300&
300&
300&
330&
480\tabularnewline
\hline
\end{tabular}\\

\begin{tabular}{cccc}
\hline 
Wavelengths $(\mu m)$&
250&
350&
550\tabularnewline
\hline 
Herschel SPIRE FWHM ('')&
17&
24&
35\tabularnewline
\hline
\end{tabular}

\end{table}

\begin{figure}
\begin{centering}\includegraphics[width=1\columnwidth]{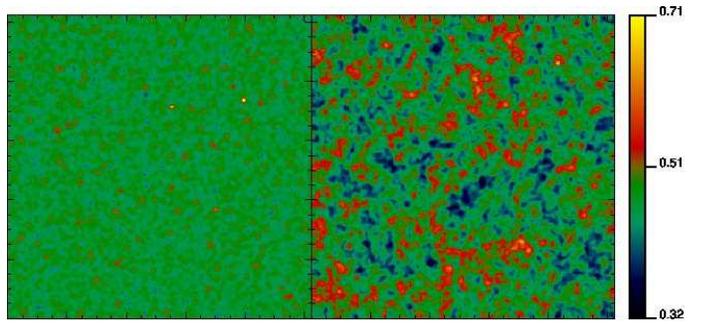}\par\end{centering}

\caption{Planck maps at 550 $\mu m$ in MJy/sr with b=0 ({\it left})
and b=1.5 ({\it right}).
The maps simulate a region of the sky of 49 square degrees with $1024$
pixels of 25 arcsec\ensuremath{}.\label{fig:Maps-Planck-550}}
\end{figure}

\begin{figure}
\includegraphics[width=1\columnwidth]{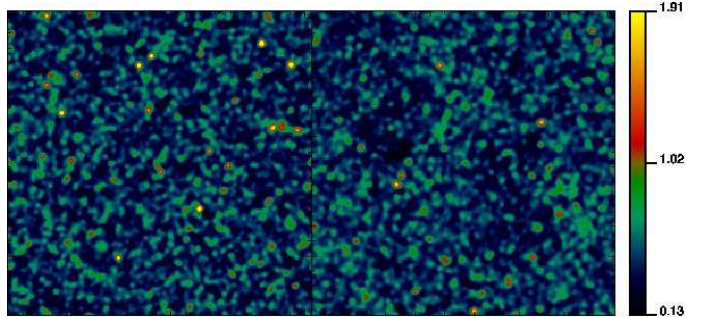}

\caption{Herschel maps at 500 $\mu m$ in MJy/sr with b=0 ({\it left}) and b=1.5
({\it right}). The maps simulate a region of the sky of 0.3 square degrees
with $1024$ pixels of 2 arcsec\ensuremath{}. The small size
of the maps makes it difficult to appreciate the effect of the large-scale clustering.\label{fig:Maps-Herschel-500}}
\end{figure}

For the purpose of this paper we have created Planck/HFI maps 
at 350, 550, 850, 1380 and 2097 microns and Herschel/SPIRE maps
at 250, 350 and 500 microns. A description of the wavelengths and
spatial resolution of the maps are given in Table \ref{cap:Wavelengths and psf of maps}.
Three different biases were used for the
simulations ($b=0, 0.75, 1.5$). Examples of maps at 500 $\mu m$
(Herschel) and 550 $\mu m$ (Planck) made with b=1.5 and b=0 are shown
in Figs. \ref{fig:Maps-Planck-550} and \ref{fig:Maps-Herschel-500}.
The difference in the spatial correlations is easily noticed in the Planck
simulations. On the other hand, the smaller size of the Herschel
simulations makes it more difficult to see the correlation.\\

The simulated maps and their associated catalogs are publicly 
available at http://www.ias.u-psud.fr/irgalaxies/simulations.php.

\section{\label{noise} NOISE AND SOURCE DETECTION}
The simulations can be used to test the
detection capabilities of Planck/HFI and Herschel/SPIRE. For the first
time these simulations use an empirical model that reproduces all
the observational constraints from 5 $\mu$m to 1.3 mm and include the spatial correlation
between the IR galaxies and the dark matter density field for galaxies up to very
low luminosities ($L>10^{9}L_{\odot}$). They provide a useful tool for 
preparing future observations with Planck/HFI and Herschel/SPIRE.

\subsection{Detection of bright sources\protect \\
}

As stated previously bright sources dominate the power spectrum of
the FIRB (see Fig. \ref{fig: Fluctuations-BackGround-S}). We therefore
need 
to subtract them before studying the fluctuations in the
background. In this section we concentrate on detecting them 
in three steps: 1) wavelet filtering, 2) detection,
3) measurement of the flux.\\

\begin{itemize}
\item Wavelet filtering: Before trying to detect the sources we perform
a wavelet transform of our simulated maps with the ``atrou'' algorithm. We remove spatial
frequencies that are both higher and lower than the FWHM of the PSF.\\

The small-scale filtering improves the estimation of the position
of the sources when the instrumental noise is included in the
simulations. 
In contrast to the confusion
noise, the instrumental noise is not correlated for neighbouring pixels.
This dominates errors in estimating the position of the sources.\\

The large-scale filtering corrects for a bias in the detection algorithm.
The algorithm searches for sources using the absolute value of the
pixel and not its value relative to its environment. This biases the
detections towards sources in bright regions. The removal of the large
spatial fluctuations corrects this effect.\\

The selection of spatial frequencies to be used for the detection has been manually optimised for each map
to achieve a maximum number of reliable detections. This treatment
is similar to what was done in the MIPS Spitzer maps \citep{2004ApJS..154...87D}.
A comprehensive study of the application of the wavelet filtering technique
for the source detections at long wavelengths for Planck and Herschel/SPIRE
is beyond the scope of this paper and has been fully discussed in
e.g. \citet{2006MNRAS.370.2047L}.\\

\item Detection algorithm: The algorithm is based on the ``find'' routine of
the DAOPHOT library. In the filtered image the algorithm searches
for peaks higher than a certain threshold $\sigma_{thres}$. It uses
the PSF shape and the neighbouring pixels to analyse whether the peak is
the centre of a source.\\

\item Flux measurement: We developed a PSF fitting algorithm that we
used in the original
map (without filtering) to measure the flux of the sources. We decided
whether the detections are real or false by two criteria: 1) proximity
and 2) accuracy (see Sect. 4.1.1).\\
\end{itemize}

\subsubsection{Bad detections}

A detection is considered good or bad based on two criteria: 1) proximity
with the position of an input source
and 2) accuracy of flux for this source. The former requires that our detection is closer
than FWHM/5 to at least one {}``neighbour'' source in our catalog.
The latter requires that the difference between the flux of one of
the {}``neighbour'' sources in the catalog and that of the detected
source has to be smaller than the confusion and/or instrumental noise
(see Tables \ref{tab:Stddev-flux-Planck} and \ref{tab:Stddev-flux-Herschel}).
We consider the detection to be good only if both criteria are satisfied.\\

\begin{figure}
\includegraphics[width=0.5\columnwidth]{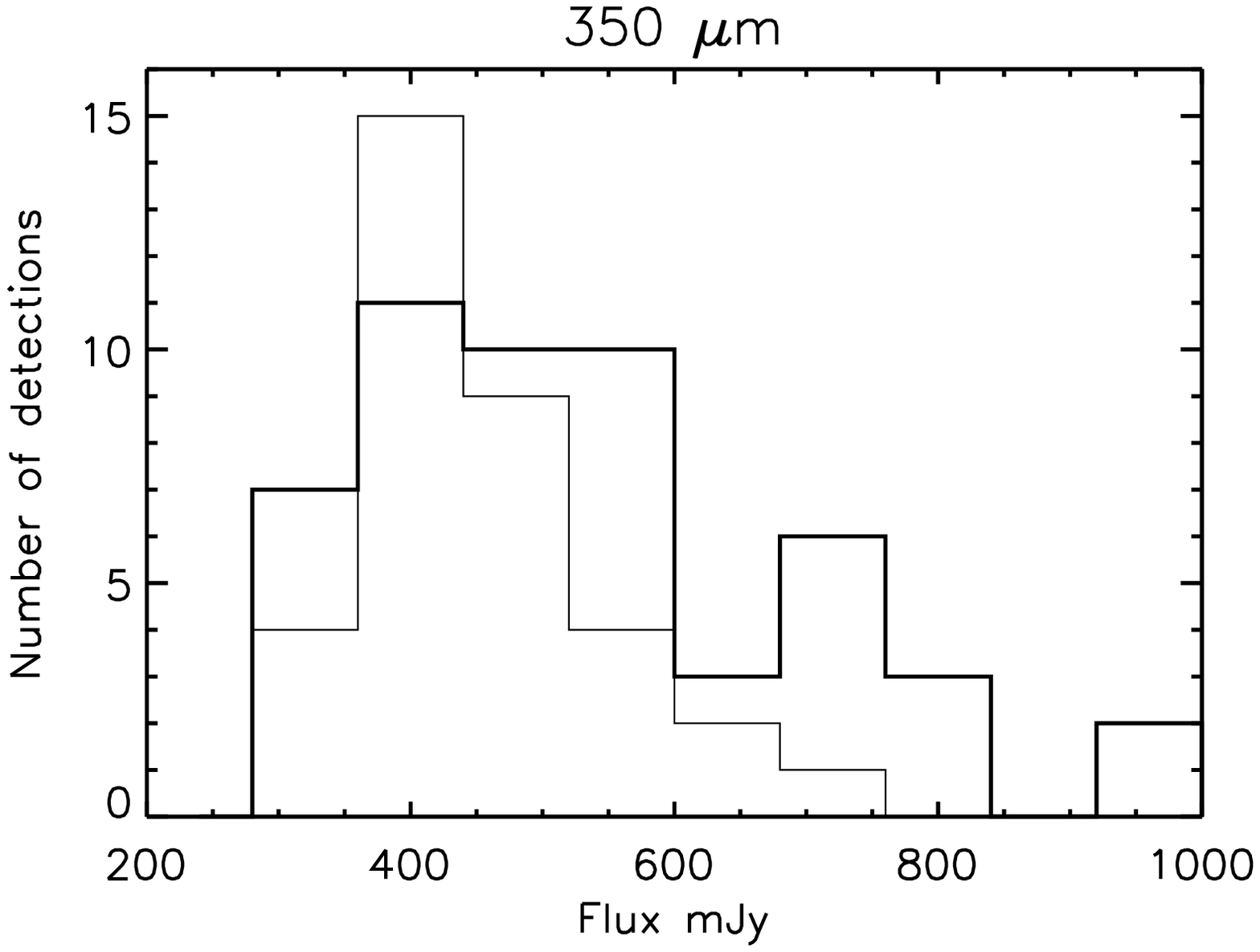}\includegraphics[width=0.5\columnwidth]{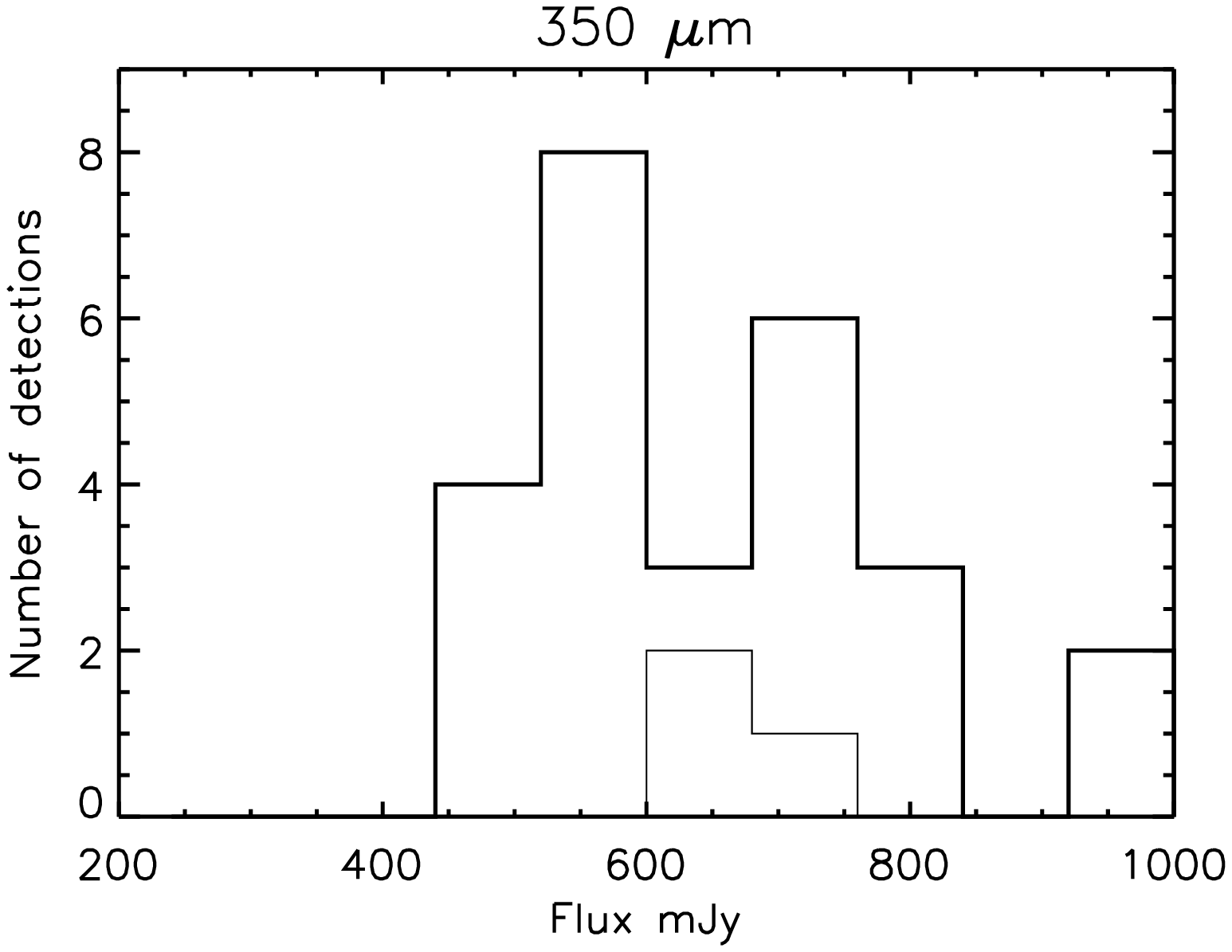}

\caption{Good detections (thick line) vs bad detections (thin line) 
{\it Left:} Histogram of good and bad detections using small $\sigma_{thres}$
(290 mJy).
{\it Right:} Histogram of good and bad detections using higher $\sigma_{thres}$
(440 mJy) in the same map. 
Both plots have been done using a Planck simulated map with $b=0$.\label{fig:Completeness-vs-Spurious}}
\end{figure}

The detection process also produces detections that do not comply
with these criteria. We can see in Fig. \ref{fig:Completeness-vs-Spurious}
how the different $\sigma_{thres}$ modify the rate of good-to-bad
detections. For a low detection threshold ($\sigma_{thres}$=$2\sigma_{map}$,
i.e. for example 290 mJy/pix at Planck 350 $\mu m$ for a map
with b=0 and no instrumental noise), 
the number of bad detections can become
bigger than that of real detections. For a higher detection threshold
($\sigma_{thres}$=$3\sigma_{map}$ i.e. 440 mJy/pix at 350 $\mu m$),
we find that the good detections dominate the bad ones, but we do not
detect as many faint sources. Thus the number of false detections
depends strongly on $\sigma_{thres}$. For different scientific goals,
it can be interesting to use different $\sigma_{thres}$. For example,
if we are interested in searching for objects at high redshifts, we
could allow our detections to have 25\% bad sources 
to be able to detect some interesting sources at high $z$. For studies
of statistical properties of the sources, it would be necessary to
use a stronger threshold. For our purpose, we used $\sigma_{thres}=3\sigma_{map}$
($\sim10\%$ of false detections).

\subsection{Instrumental and confusion noises}

Instrumental and confusion noises have been studied both separately
and in combination in order to quantify their relative contribution
to the total noise. The estimated instrumental noises
per beam for Planck and Herschel are given in Table \ref{tab:Instrumental-noise-mJy}.
The instrumental noise per beam for Planck is the average one over the
sky  
for a 1-year mission. The instrumental noise per beam for Herschel
is typical of large surveys. We take the sensitivity of the so-called
level 5 and level 6 of the Science Activity Group 1 (SAG 1) of the
SPIRE guaranteed time team.\\

We studied the standard deviation of the measured fluxes in random
positions for different maps. These maps were one of instrumental
noise, three with different bias (b=0, 0.75, 1.5) but without instrumental noise, 
and complete maps created
by adding the map of instrumental noise to the three source maps.
We  call these maps hereafter instrumental-only, confusion-only,
and complete-maps. We fit a Gaussian to the histogram of the fluxes
measured in these random positions and considered the standard deviation
of this Gaussian as the best estimate of the standard deviation of
the photometry of a source and therefore of the 1$\sigma$ instrumental, confusion,
and total noise. Results are shown in Tables \ref{tab:Stddev-flux-Planck}
and \ref{tab:Stddev-flux-Herschel}. \\

The confusion noise increases with the bias. This effect is noticeable
for the Planck observations, but not for the Herschel ones because
of the higher Herschel/SPIRE angular resolution. Also, for the
considered Herschel/SPIRE
surveys, the instrumental noise is always greater than the confusion
noise. 
For Planck the correlation effect is more noticeable
for longer wavelengths since they probe progressively higher redshifts
and therefore higher 
dark-matter power spectra, as discussed in Sect 2.3 (see Fig. \ref{fig:dCl/dz}).\\

The total noise $\sigma_{C+I}$ is close to the 
value $\sigma_{C+I}^{2}=\sigma_{C}^{2}+\sigma_{I}^{2}$ (see Tables 
\ref{tab:Stddev-flux-Planck} and \ref{tab:Stddev-flux-Herschel}). For Planck
at short wavelengths (350 $\mu m$ and 550 $\mu m$), the confusion
noise is the dominant source of noise. The instrumental noise becomes
dominant at $850\mu m$ for b=0 and b=0.75. For longer wavelengths,
it dominates for any bias. For Herschel the instrumental noise dominates
the total noise for both the shallow and deep surveys. The confusion
noise is not strongly affected by the bias because of the small FWHM of the
PSF.\\

\begin{table}

\caption{Simulation input instrumental noise per pixel of size equal 
to beam for Planck and for Herschel for a deep
and a shallow survey \label{tab:Instrumental-noise-mJy}.}

\begin{tabular}{cccccc}
\hline 
Wavelengths HFI $(\mu m)$&
350&
550&
850&
1380&
2097\tabularnewline
\hline 
$\sigma_{Inst}$(mJy)&
31.30&
20.06&
14.07&
8.43&
6.38\tabularnewline
\hline
\end{tabular}\\

\begin{tabular}{cccc}
\hline 
Wavelengths SPIRE $(\mu m)$&
250&
350&
500\tabularnewline
\hline 
$\sigma_{Inst}$ Deep (mJy)&
4.5&
6.1&
5.3\tabularnewline
$\sigma_{Inst}$ Shallow (mJy)&
7.8&
10.5&
9.2\tabularnewline
\hline
\end{tabular}\\

\end{table}

\begin{table}

\caption{Noise on the retrieved sources with only instrumental noise ($\sigma_{I}$), confusion noise ($\sigma_{C}$),
and total noise ($\sigma_{C+I}$) in mJy for Planck/HFI. \label{tab:Stddev-flux-Planck}}

\begin{tabular}{cccccc}
\hline

Wavelengths HFI ($\mu m$)&
350&
550&
850&
1380&
2097\tabularnewline
\hline 
$\sigma_{I}$ &
61.3&
39.2&
27.8&
16.7&
12.5\tabularnewline
$\sigma_{C}$ b=0&
111.5&
41.5&
14.7&
4.6&
2.1\tabularnewline
$\sigma_{C}$ b=0.75&
124&
54.3&
21.3&
7.7&
3.5\tabularnewline
$\sigma_{C}$ b=1.5&
158&
79.8&
30.4&
10.9&
5.4\tabularnewline
$\sigma_{C+I}$ b=0&
126.7&
62.3&
33&
17.4&
13.2\tabularnewline
$\sigma_{C+I}$ b=0.75&
153.6&
75.3&
38.4&
19&
13.8\tabularnewline
$\sigma_{C+I}$ b=1.5&
188.2&
95.3&
46.7&
21.1&
14.4\tabularnewline
\hline
\end{tabular}

\end{table}

\begin{table}

\caption{Instrumental noise ($\sigma_{I}$), confusion noise ($\sigma_{C}$)
and total noise ($\sigma_{C+I}$) in mJy for Herschel/SPIRE. \label{tab:Stddev-flux-Herschel}}

\begin{tabular}{cccc}
\hline 
Wavelengths SPIRE $(\mu m)$&
250&
350&
500\tabularnewline
\hline 
Deep $\sigma_{I}$&
8.7&
11.3&
10.1\tabularnewline
Shallow $\sigma_{I}$&
15&
20.1&
18.2\tabularnewline
$\sigma_{C}$ b=0, 0.75, 1.5&
4.6&
6.5&
5.5\tabularnewline
Deep $\sigma_{C+I}$ &
9.8&
12.3&
11\tabularnewline
Shallow $\sigma_{C+I}$ &
16&
20.8&
19\tabularnewline
\hline
\end{tabular}

\end{table}

\subsection{Completeness}

The first study of the point-sources detection limit for Planck was
carried out based on a generalisation of the Wiener filtering method
\citep{bouchet-1999-4}. Recently \citet{2006MNRAS.370.2047L}
used the most recent available templates of the microwave sky
and extragalactic point source simulations, including both the radio
and IR galaxies, to estimate the Planck
detection limits. Here we revisit those results with new models for
IR galaxies and the last noise estimates for Planck/HFI and
Herschel/SPIRE.\\

\begin{figure}
\includegraphics[width=1\columnwidth]{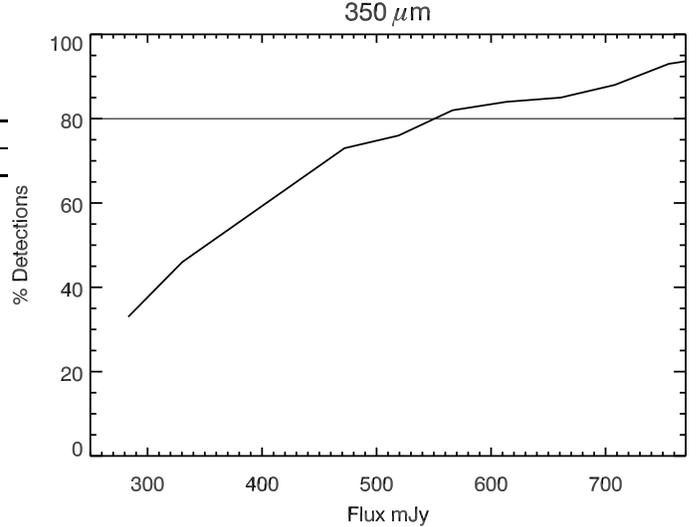}

\caption{Study of completeness for Planck/HFI with $b=0$ at $350\mu m$. The horizontal
straight line marks 80\% of completeness. \label{fig:Study-completeness}}
\end{figure}

For the study of the completeness, a number $N_{A}$
of sources of equal flux are randomnly distributed in the maps.
Each source is placed far enough from the other to avoid  these
additional sources contributing 
to the confusion noise. The detection
and photometry of these sources is carried out as described at the
beginning of the section. We call $N_{G}$ the number of good detections
that comply with the {}``proximity'' and {}``accuracy'' criteria.
The completeness for this flux $C_{F}$ is then calculated as $C_{F}=\frac{N_{G}}{N_{A}}\times100$.
The completeness of the detections of sources for a given flux depends
on both the instrumental noise and the confusion noises. The results
for the completeness are averaged over $\sim$3000 individual
fake sources per flux. An example of the completness at 350$\mu m$ is
shown in Fig. \ref{fig:Study-completeness}.
Results for all wavelengths
are given in Tables \ref{tab:Completeness-Planck} and \ref{tab:Completeness-Herschel}.
They are
consistent with the instrumental, confusion, and total noises given in Sect.
4.2 and the conclusions from that section remain valid for the completeness.
For simulated maps including both extragalactic sources and
instrumental noise, we find that the 80\% completeness level coincide
with flux limits around $4-5\sigma$.\\

\begin{table}

\caption{\label{tab:Completeness-Planck}Completeness limits (in mJy) for the Planck/HFI maps
with instrumental noise ($C_{I}$), confusion noise ($C_{C}$) and
both ($C_{C+I}$). We consider $b$=0, 0.75, and 1.5.}

\begin{tabular}{cccccc}
\hline 
Wavelengths HFI ($\mu m$)&
350&
550&
850&
1380&
2097\tabularnewline
\hline 
$C_{I}=80\%$ &
236&
157&
108&
67&
50\tabularnewline
$C_{C}=80\%$ b=0&
516&
174&
60.5&
20&
8.6\tabularnewline
$C_{C}=80\%$ b=0.75&
550&
239.5&
88.5&
30.5&
15.5\tabularnewline
$C_{C}=80\%$ b=1.5&
684&
300&
121&
40&
24\tabularnewline
$C_{C+I}=80\%$ b=0&
560&
234&
126&
71&
52\tabularnewline
$C_{C+I}=80\%$b=0.75&
607&
290&
141&
74&
55\tabularnewline
$C_{C+I}=80\%$ b=1.5&
709&
360&
171&
80&
58.5\tabularnewline
\hline
\end{tabular}

\end{table}

\begin{table}

\caption{\label{tab:Completeness-Herschel}Completeness limits (in mJy) for the Herschel/SPIRE
maps with instrumental noise ($C_{I}$), confusion noise ($C_{C}$),
and both ($C_{C+I}$).}

\begin{tabular}{cccc}
\hline 
Wavelengths SPIRE$(\mu m)$&
250&
350&
500\tabularnewline
\hline 
Deep $C_{I}=80\%$&
33&
45.9&
37.9\tabularnewline
Shallow $C_{I}=80\%$&
57.3&
84.2&
67.3\tabularnewline
$C_{C}=80\%$&
35&
32&
27.4\tabularnewline
Deep $C_{C+I}=80\%$&
49.8&
64.4&
48.4\tabularnewline
Shallow $C_{C+I}=80\%$&
70&
96.5&
75.5\tabularnewline
\hline
\end{tabular}

\end{table}

Taking these 80\% completness limits as a detection threshold,
the prediciton for the number of sources detected directly by
Herschel/SPIRE for the deeper survey considered here
is 8.3$\times10^{5}/sr$ at 250 $\mu m$, 1.1$\times 10^{5}/sr$ at 350 $\mu m$, and
1.8$\times 10^{4}/sr$ at 500 $\mu m$. The fraction of resolved CFIRB varies
between 8 and 0.3\% from 250 to 500 $\mu m$. \\

\section{CIB fluctuations}\label{fluctuations}
The Planck and Herschel/SPIRE surveys allow an unprecedented search for CFIRB fluctuations 
associated with large-scale structure and galaxy clustering. Background fluctuations probe the physics 
of galaxy clustering over an ensemble of sources, with the bulk of the signal contribution originating 
from sources well below the detection threshold. Thus a comprehensive fluctuation 
analysis is an essential complement to the study of individually
detected galaxies.
In this section, we restrict ourselves to predictions for Planck/HFI,
excluding the 143 and 100 GHz channels. At these low frequencies,
we are dominating by the non-thermal emission of the
radiosources -- that are not including in our model -- and 
the Poisson term dominates the clustering term
\citep [e.g.] [] {2005ApJ...621....1G}. Also, we exclude the
Herschel/SPIRE case since our simulations include the
clustering of CIB sources in two different halos ($2h$), but not the clustering
within the same halo ($1h$). The $1h$ term dominates for $\ell \gtrsim
3000$ and will be accurately measured by Herschel/SPIRE. Only
large-scale surveys can put strong constraints on the $2h$ term.
Measuring the $2h$ clustering with CFIRB anisotropies is one of the
goals of Planck/HFI.

\subsection{Contributors to the angular power spectrum}

From the far-IR to the millimeter, the sky is made up of the CFIRB and two other sources of
signal, the galactic cirrus and the CMB (we neglect the SZ signal). Understanding our observations
of the CFIRB requires understanding the contributions from these two
components which act for us as foreground and background contamination.\\

The galactic cirrus acts as  foreground noise for the CFIRB. The non-white
and non-Gaussian statistical properties of its emission make it a
very complex foreground component. The power spectrum of
the IRAS 100 $\mu m$ emission is characterised by a power law \citet
[e.g.] [] {1992AJ....103.1313G}.
Here we compute the angular power spectrum
of the dust emission following \citet{2007arXiv0704.2175M}.
These authors analysed the statistical properties of the cirrus
emission at 100 $\mu$m using the IRAS/IRIS data.
We used their power spectrum normalization and slope (varying with the
mean dust intensity at 100$\mu$m). Using the average $|b|>30^o$ spectrum of the HI-correlated
dust emission measured using FIRAS data,
we converted the 100 $\mu$m power spectra to the Planck
wavelengths. For the discussion, we considered both the total cirrus
fluctuations and 10\% residual fluctuations. These 10\% could be
achievable with Planck in low dust-column-density regions containing
ancillary HI data.\\

The CMB acts as background noise for the CFIRB. However, the CMB angular
power spectrum is known to an accuracy of 1\% or better (Planck-HFI web site:
http://www.planck.fr/). This combines with its well-known spectral
dependence to allow for a clean subtraction of its contribution that
in turn allows
for detections of the CFIRB $C_{l}$ even for wavelengths where the
CMB dominates. We consider for the rest of the discussion a
conservative assumption, that is that 
the residual CMB fluctuations are approximately 2\%.\\

The angular power spectrum of IR galaxies is composed of a correlated
and a Poissonian part. As discussed in Sect. 2.1, the contribution
to the Poissonian part is dominated by relatively faint sources after
subtracting the brightest galaxies (see Fig. \ref{fig: Fluctuations-BackGround-S}).
We consider that we can remove sources brighter than our
80\% completeness detection limit (see Tables \ref{tab:Completeness-Planck} and
\ref{tab:Completeness-Herschel}). For doing so, we use the technique
described in Sect. 4.1 for measuring the position and fluxes of the
sources and once these are known we subtract a PSF with the measured
flux from the map.\\

The correlated $C_{l}$ is obtained as described in Sect 2.1. For
the relative error on the power spectrum, we follow \citet{1995PhRvD..52.4307K}:

\[
\frac{\delta C_{l}}{C_{l}}=\left(\frac{4\pi}{A}\right)^{0.5}\left(\frac{2}{2l+1}\right)^{0.5}\left(1+\frac{A\sigma_{pix}^{2}}{NC_{l}W_{l}}\right)\]

where A is the observed area, $\sigma_{pix}$ the rms noise per
pixel (instrumental plus confusion), N the number of pixels, and $W_{l}$ the window function for
a map made with a Gaussian beam $W_{l}=e^{-l^{2}\sigma_{B}^{2}}$.\\
\begin{figure*}[tp]
\begin{centering}\includegraphics[width=0.85\columnwidth,keepaspectratio]{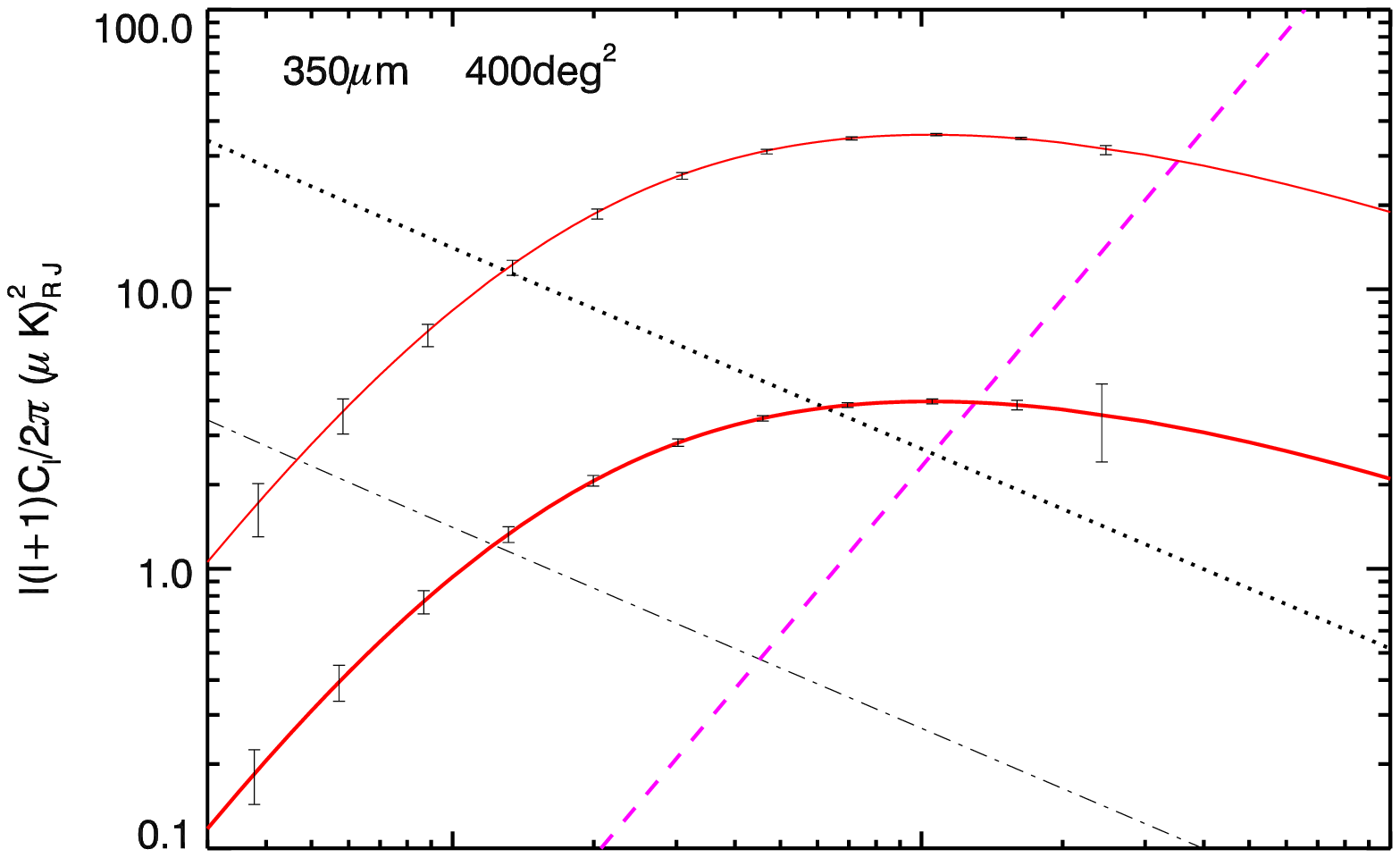}\includegraphics[width=0.73\columnwidth,keepaspectratio]{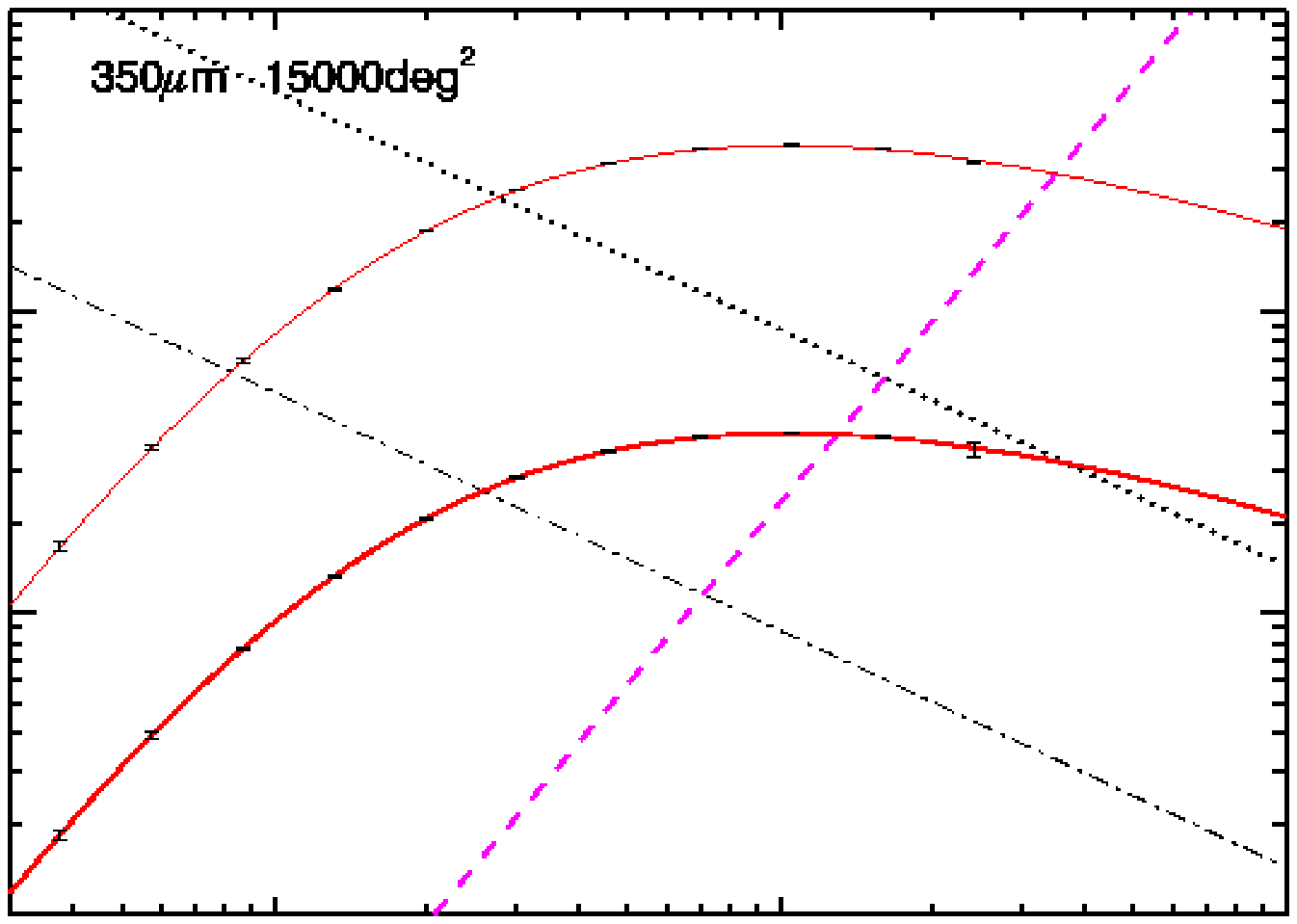}\par\end{centering}

\begin{centering}\includegraphics[width=0.85\columnwidth,keepaspectratio]{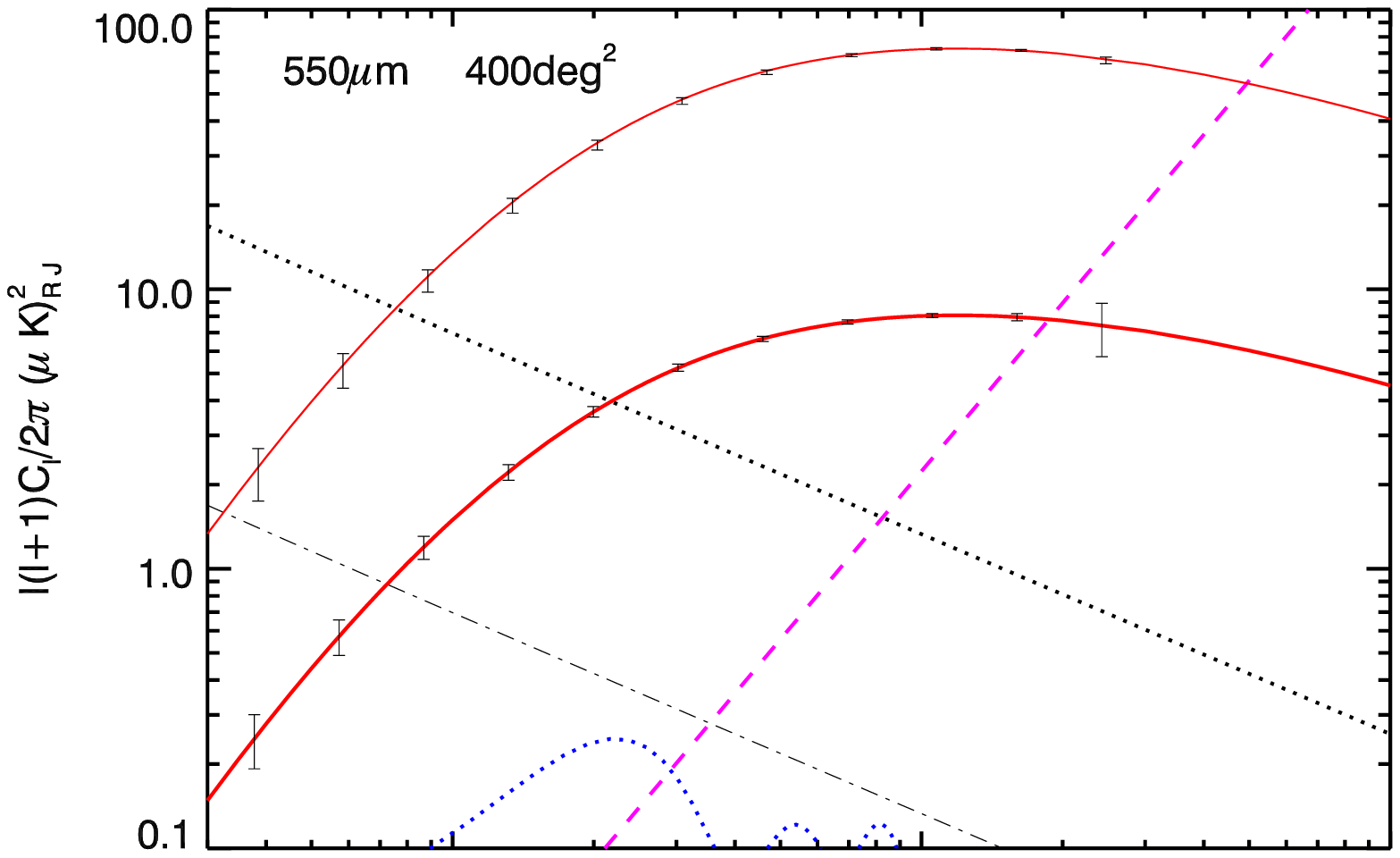}\includegraphics[width=0.73\columnwidth,keepaspectratio]{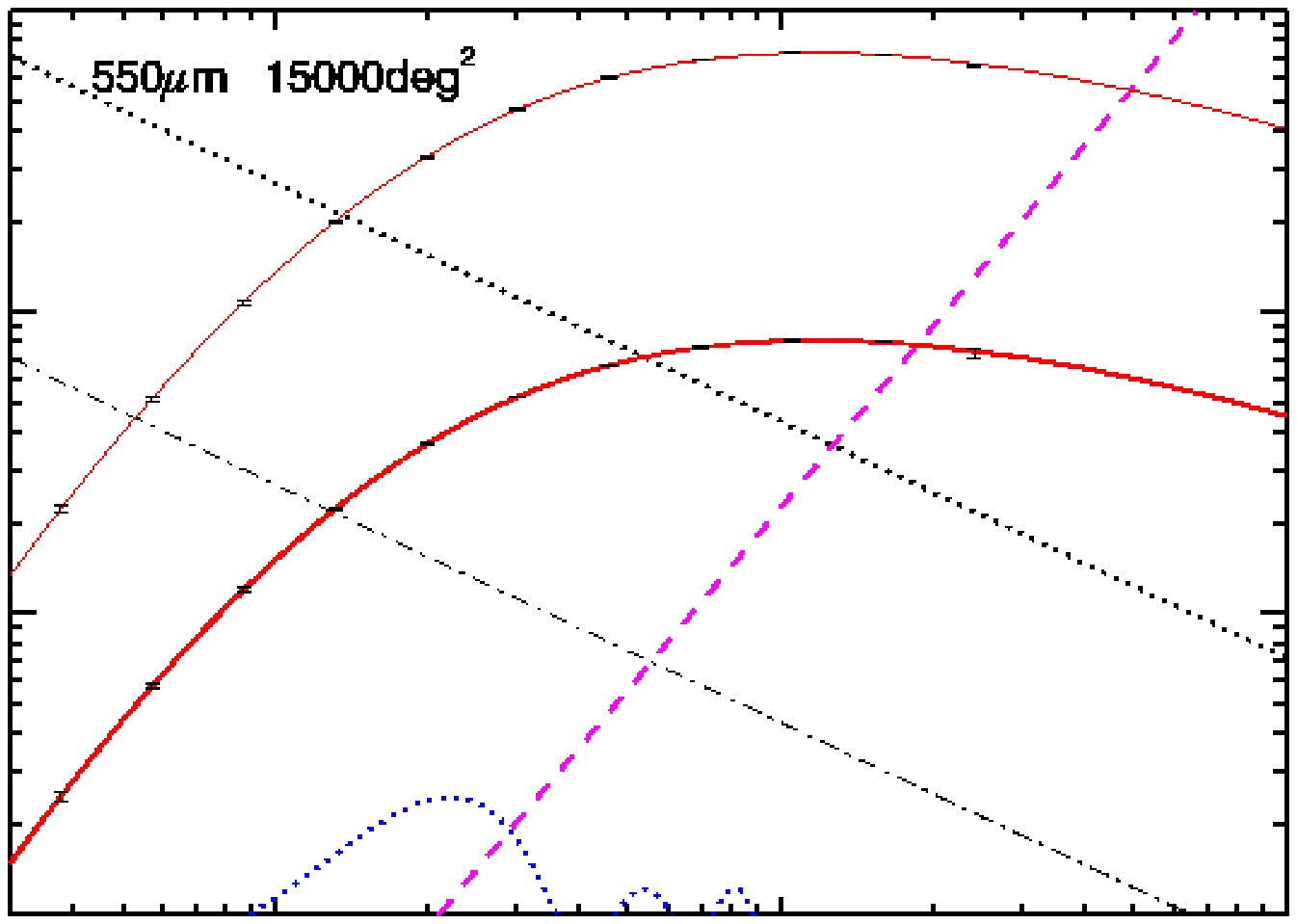}\par\end{centering}

\begin{centering}\includegraphics[width=0.85\columnwidth,keepaspectratio]{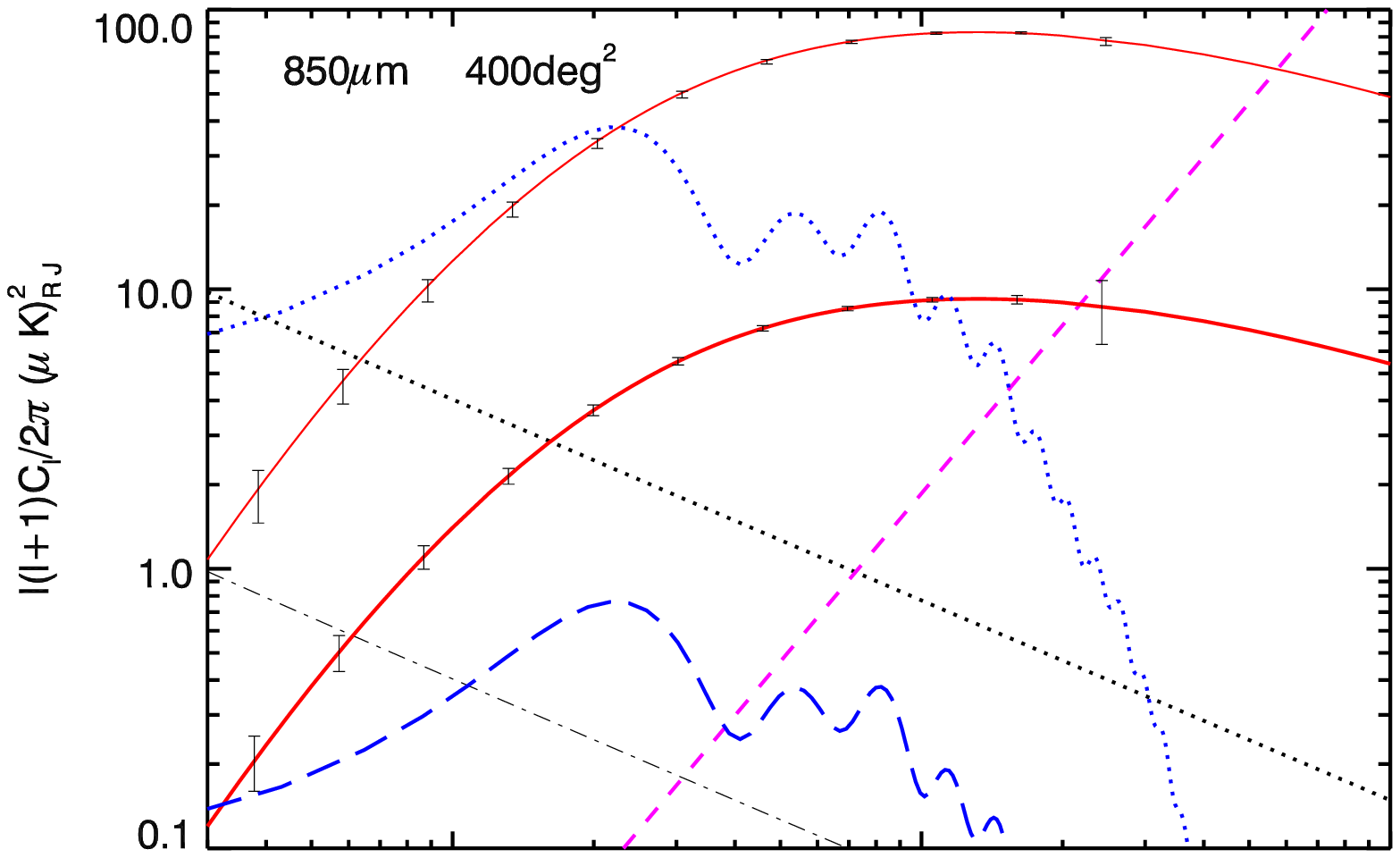}\includegraphics[width=0.73\columnwidth]{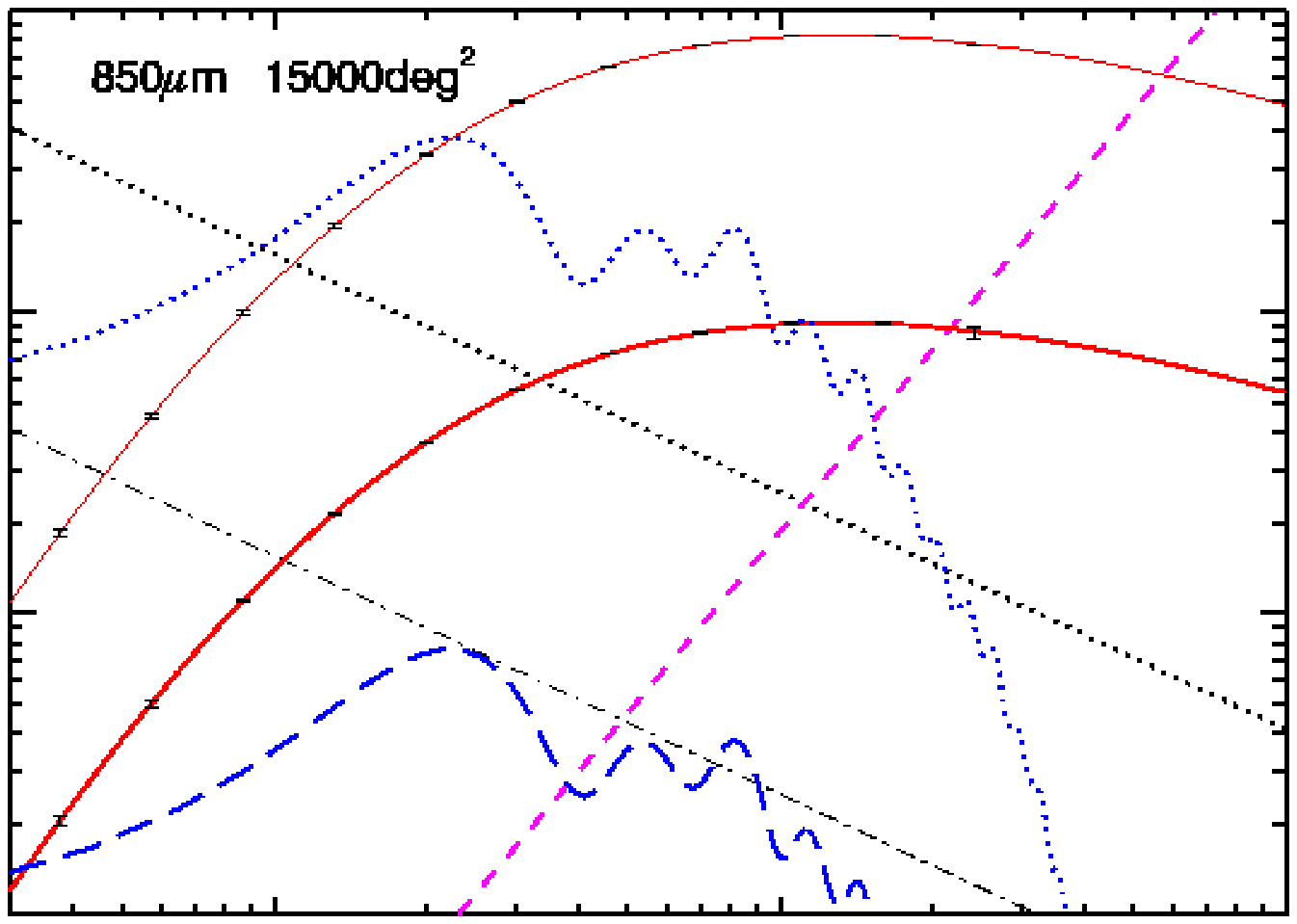}\par\end{centering}

\begin{centering}\includegraphics[width=0.85\columnwidth,keepaspectratio]{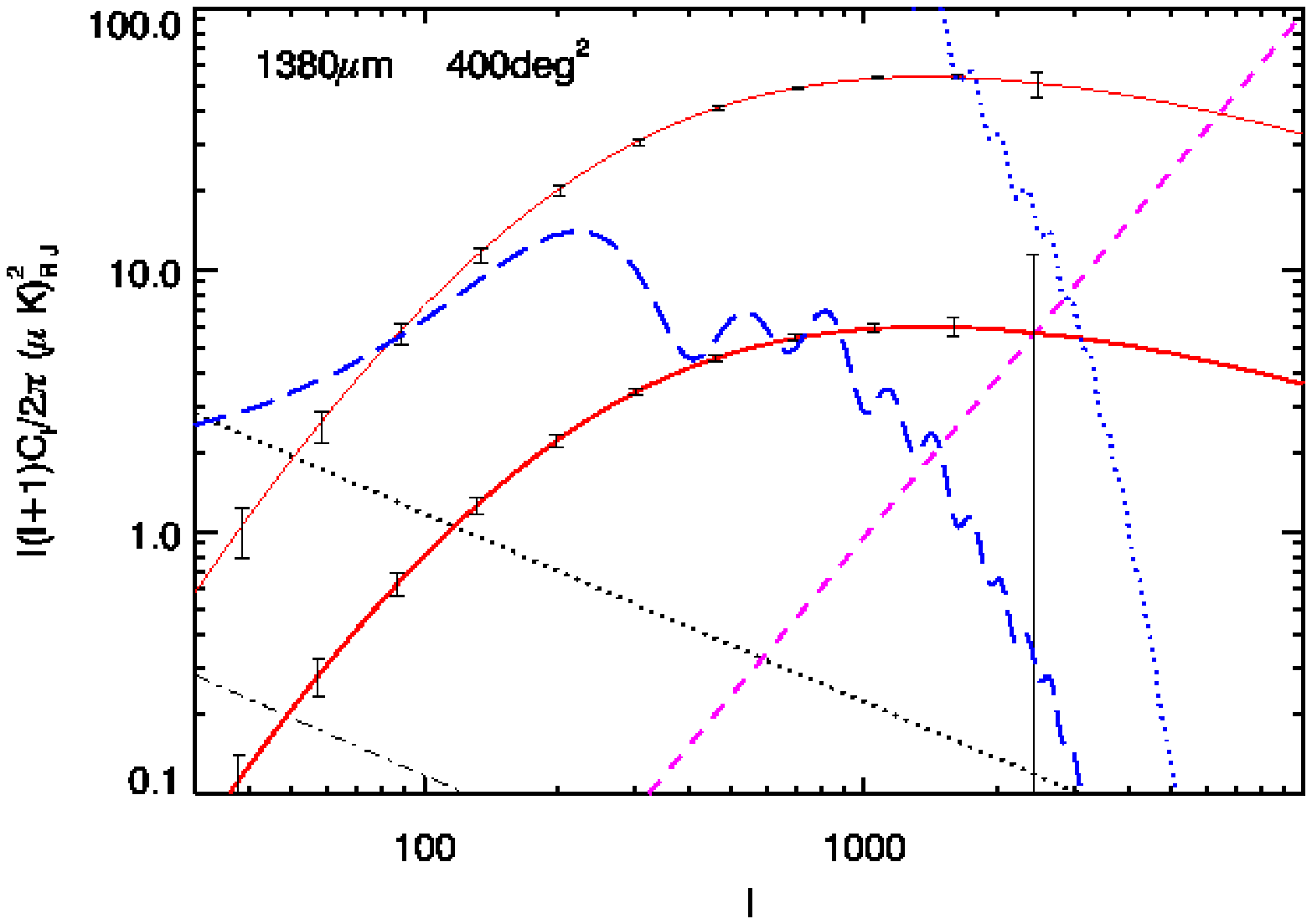}\includegraphics[width=0.73\columnwidth]{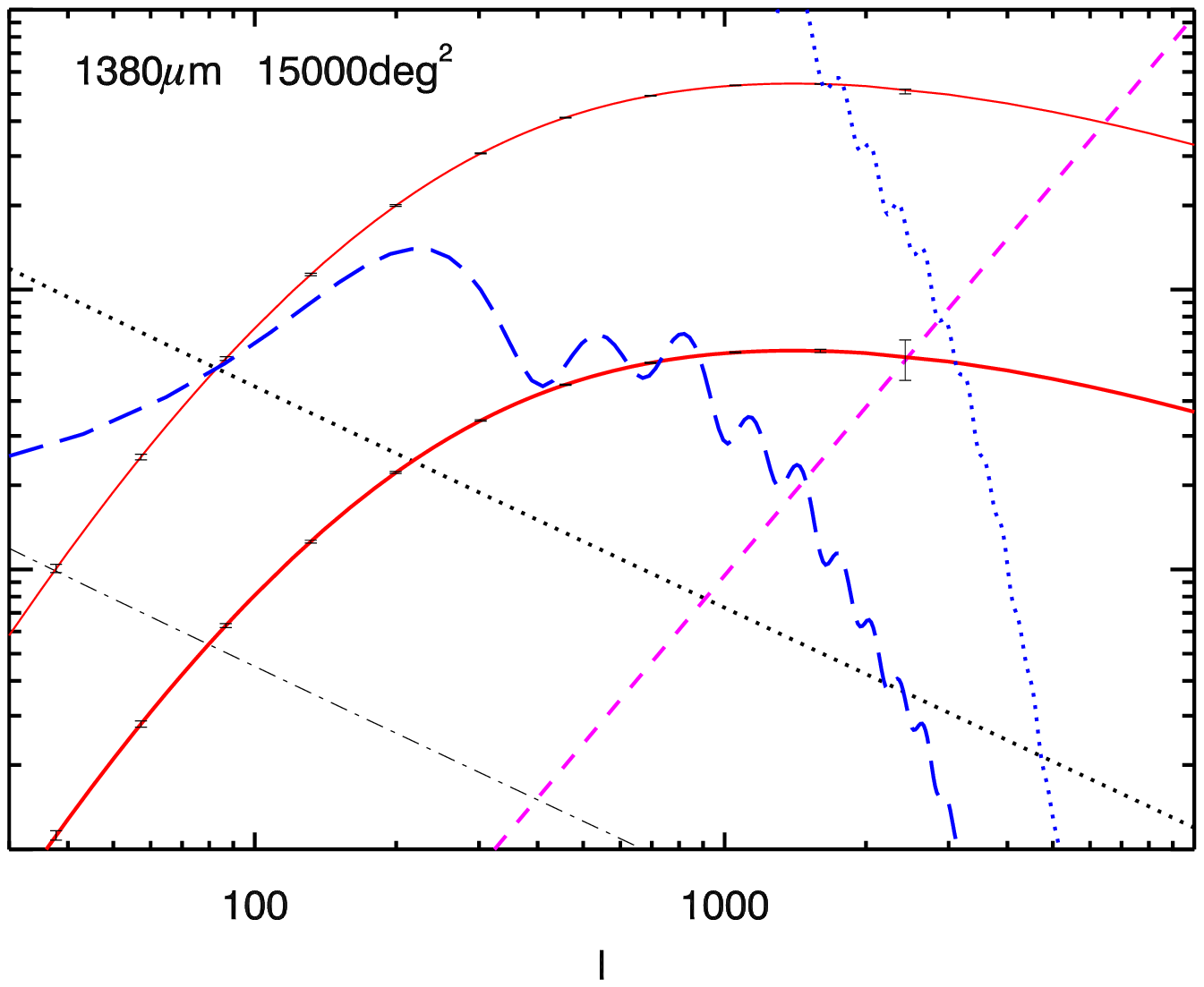}\par\end{centering}

\caption{Power spectra for Planck of the correlated IR galaxies with b=1 (thick
continuous line) and b=3 (thin continuous line), Poisson fluctuations
for sources fainter than 709, 363, 171, 80 mJy at 350, 550, 850 and
1380 $\mu m$ respectively (short dashed line), dust with an HI column density
of 1.5 $10^{20}$ $at/cm^{2}$ and 2.7 $10^{20}$ $at/cm^{2}$ for
the 400 Sq. Deg. and $\sim$15000 Sq. Deg. considered regions respectively (dotted line), dust/10
(dotted-dashed line), CMB (dotted line) and CMB/50 (dashed line). Error bars are computed using 
$\Delta \ell / \ell $=0.5.
\label{fig:Power-spectra-Planck}}
\end{figure*}

\subsection{Detectability of CFIRB correlated anisotropies}
The study of the $C_{l}$ on different scales allows us to study different
aspects of the physics of the environment of IR galaxies \citep [see] [] {2002PhR...372....1C}.
Large scales ($l<100$) give information on the cosmological evolution
of primordial density fluctuations in the linear phase and therefore
on the cosmological parameters. Intermediate scales are mostly influenced
by the mass of dark halos hosting sources, which determines the bias
parameter. Small scales ($l>3000$) probe the distribution of sources
in the dark-matter halos and therefore the non-linear evolution of
the structures. This non linear evolution was not accounted for in
our model.\\

\begin{table*}

\caption{Multipole $\ell$ ranges where the CFIRB correlated
anisotropies are higher than
the cirrus and CMB components.\label{tab:Cl-Range-Planck} }

\begin{tabular}{ccccc}
\hline 
Wavelengths ($\mu m$)&
350&
550&
850&
1380\tabularnewline
\hline 
400 Sq. Deg.\ensuremath{} b=1&
 $600<\ell<1200$&
 $200<\ell<1800$&
 $1000<\ell<2200$&
Indetectable\tabularnewline
400 Sq. Deg.\ensuremath{} b=3&
 $120<\ell<3500$&
 $75<\ell<5000$&
 $240<\ell<6000$&
 $1800<\ell<6500$\tabularnewline
15000 Sq. Deg.\ensuremath{} b=1&
Indetectable&
 $550<\ell<1800$&
 $1200<\ell<2200$&
Indetectable\tabularnewline
15000 Sq. Deg.\ensuremath{} b=3&
 $280<\ell<3500$&
 $130<\ell<5000$&
 $230<\ell<6000$&
 $1800<\ell<6500$\tabularnewline
\hline
\end{tabular}

\end{table*}

\begin{table*}

\caption{Multipole $\ell$ ranges where the CFIRB correlated anisotropies
are higher than
the residual cirrus and CMB components (10\%, and 2\% respectively). \label{tab:Cl-Range-Planck-Subtracted}}

\begin{tabular}{cccccc}
\hline 
Wavelengths ($\mu m$)&
350&
550&
850&
1380\tabularnewline
\hline 
400 Sq. Deg.\ensuremath{} b=1&
$200<l<1200$&
 $80<l<2100$&
 $60<l<2200$&
Indetectable\tabularnewline
400 Sq. Deg.\ensuremath{} b=3&
 $40<l<3500$&
 $30<l<5000$&
 $25<l<6000$&
 $80<l<6500$\tabularnewline
15000 Sq. Deg.\ensuremath{} b=1&
 $250<l<1200$&
 $130<l<1800$&
 $100<l<2200$&
Indetectable\tabularnewline
15000 Sq. Deg.\ensuremath{} b=3&
 $80<l<3500$&
 $55<l<5000$&
 $45<l<6000$&
 $80<l<6500$\tabularnewline
\hline
\end{tabular}

\end{table*}

We can see in Fig. \ref{fig:Power-spectra-Planck} 
the different contributions to the $C_{l}$ for Planck/HFI: 
the correlated CFIRB $C_{l}$ with $b=1$ and $b=3$ with their respective
error bars ($\Delta \ell / \ell $=0.5), Poissonian fluctuations, dust, and CMB contributions.
The dust and CMB $C_{l}$ are plotted both before and
after implementing the corrections discussed above (10\% and 2\%
residuals, respectively). For the dust we
selected for the estimation of the column density a field
centred on the SWIRE/ELAIS S1 field. This field has a very low level
of dust contamination. For a 400 sq. deg. area centred at
$(l,b)=(311^o,-73^o)$, we have an average 
HI column density of only 1.5 10$^{20}$ at/cm$^{2}$.
We also take the whole sky above $|b|>40^o$  ($\sim150000$).
The average HI column density is 2.7 10$^{20}$ at/cm$^{2}$.\\

At all wavelengths, the correlated CFIRB $C_{l}$ is dominated by the other
components on both very large and very small
scales. The angular frequencies where the $C_{l}$ is dominated by
the galaxy correlation depends mainly on the wavelength, if the bias
and cirrus level are fixed. On large scales the $C_{l}^{CFIRB}$ is
dominated by the dust for wavelengths up to 550 $\mu m$, and for longer
wavelengths it is dominated by the CMB. On small scales the Poissonian
fluctuations dominate the power spectra. \\

We can see in Tables \ref{tab:Cl-Range-Planck} and \ref{tab:Cl-Range-Planck-Subtracted} 
the ranges for which CFIRB-correlated anisotropies are dominating depending on whether we consider 
a partial subtraction of the cirrus and CMB or not.
On small angular scales (large $\ell$), the range
of $\ell$ where the correlated $C_{l}$ dominates increases with the
wavelength. On large angular 
scales (small $\ell$), the results depend on whether we consider partial subtraction of the cirrus 
and CMB or not.\\

Table \ref{tab:Cl-Range-Planck} shows the $\ell$ ranges for where
CFIRB correlated anisotropies 
 are dominating when no subtraction of the dust or
CMB has been performed. 
At large angular scales and wavelengths up to 850 $\mu m$ the detectability is
better for the small map because of its lower N(HI) column density.
Without any dust contamination correction,
this prevents us from taking advantage from the smaller errror on the estimation
of the $C_l$ in the very large area surveys and makes both kind of observation
complementary. In clean regions of the sky, the best wavelengths for the observation of the large
scales CFIRB anisotropies are 350 $\mu m$ and 550 $\mu m$. \\

The large-scale detectability drastically improves in the case of
partial subtraction of the CMB and dust cirrus as seen in Table 
\ref{tab:Cl-Range-Planck-Subtracted}.
As discussed above we expect to be able to subtract the dust fluctuations
and the CMB to a 10\% and 2\% of their original level. This does not
change our ability to measure the correlated CFIRB $C_{l}$
on small scales (due to the high confusion noise),
but will allow us to probe larger scales.

\section{Summary}

This paper presented new simulations of the cosmic infrared and
submillimeter background. The simulations are based on an empirical
model of IR galaxy evolution (the LDP model) combined with a simple
description of the correlation. The IR galaxy spatial distribution follows that of
the dark-matter halo density field, with a bias parameter accounting
for the possibility of the luminous matter being more or less
correlated with the dark matter. The simulated maps and their catalogs are publicly
available
at http://www.ias.u-psud.fr/irgalaxies/simulations.php. Other maps are
available upon request. These maps are intended to be a useful tool for planning
future large IR and submillimeter surveys. In this paper, we used
the maps to predict the confusion noise and completeness
levels for Planck/HFI and Herschel/SPIRE. We also predicted the power
spectra of correlated CFIRB anisotropies for Planck/HFI which will be 
a major advance in the study of the
CFIRB anisotropies at large scales (i.e. $\ell<2000-5000$ depending
on the wavelengths). Further analysis of the CFIRB
anisotropies, including the use of stacking analysis to
isolate the anisotropies in different redshift ranges,
will be presented in a second paper, now in preparation.

\bibliographystyle{natbib}
\bibliography{8188.bbl}

\begin{thebibliography}{{Miville-Desch{\^e}nes} et~al.(2001)}

\bibitem[{Adelberger} et~al.(1998)]{1998ApJ...505...18A}
{Adelberger}, K.~L., {Steidel}, C.~C., {Giavalisco}, M., {Dickinson}, M.,
  {Pettini}, M. \& {Kellogg}, M.
\newblock September 1998, \apj, 505, 18--24.

\bibitem[{Bardeen} et~al.(1986)]{1986ApJ...304...15B}
{Bardeen}, J.~M., {Bond}, J.~R., {Kaiser}, N. \& {Szalay}, A.~S.
\newblock May 1986, \apj, 304, 15--61.

\bibitem[{Bertoldi} et~al.(2000)]{2000astro.ph.10553B}
{Bertoldi}, F., {Menten}, K.~M., {Kreysa}, E., {Carilli}, C.~L. \& {Owen}, F.
\newblock October 2000, ArXiv Astrophysics e-prints.

\bibitem[{Blain} et~al.(2004)]{2004ApJ...611..725B}
{Blain}, A.~W., {Chapman}, S.~C., {Smail}, I. \& {Ivison}, R.
\newblock August 2004, \apj, 611, 725--731.

\bibitem[Bouchet \& Gispert(1999)]{bouchet-1999-4}
Bouchet, F.~R. \& Gispert, R.
\newblock 1999, NEW ASTRON., 4, 443.

\bibitem[{Brand} et~al.(2003)]{2003NewAR..47..325B}
{Brand}, K., {Rawlings}, S., {Hill}, G.~J. \& {Lacy}, M.
\newblock September 2003, New Astronomy Review, 47, 325--328.

\bibitem[{Cooray} \& {Sheth}(2002)]{2002PhR...372....1C}
{Cooray}, A. \& {Sheth}, R.
\newblock December 2002, \physrep, 372, 1--129.

\bibitem[{Dole} et~al.(2001)]{2001A&A...372..364D}
{Dole}, H., {Gispert}, R., {Lagache}, G., {Puget}, J.-L., {Bouchet}, F.~R.,
  {Cesarsky}, C., {Ciliegi}, P., {Clements}, D.~L., {Dennefeld}, M.,
  {D{\'e}sert}, F.-X., {Elbaz}, D., {Franceschini}, A., {Guiderdoni}, B.,
  {Harwit}, M., {Lemke}, D., {Moorwood}, A.~F.~M., {Oliver}, S., {Reach},
  W.~T., {Rowan-Robinson}, M. \& {Stickel}, M.
\newblock June 2001, \aap, 372, 364--376.

\bibitem[{Dole} et~al.(2003)]{2003ApJ...585..617D}
{Dole}, H., {Lagache}, G. \& {Puget}, J.-L.
\newblock March 2003, \apj, 585, 617--629.

\bibitem[{Dole} et~al.(2004)]{2004ApJS..154...87D}
{Dole}, H., {Le Floc'h}, E., {P{\'e}rez-Gonz{\'a}lez}, P.~G., {Papovich}, C.,
  {Egami}, E., {Lagache}, G., {Alonso-Herrero}, A., {Engelbracht}, C.~W.,
  {Gordon}, K.~D., {Hines}, D.~C., {Krause}, O., {Misselt}, K.~A., {Morrison},
  J.~E., {Rieke}, G.~H., {Rieke}, M.~J., {Rigby}, J.~R., {Young}, E.~T., {Bai},
  L., {Blaylock}, M., {Neugebauer}, G., {Beichman}, C.~A., {Frayer}, D.~T.,
  {Mould}, J.~R. \& {Richards}, P.~L.
\newblock September 2004, \apjs, 154, 87--92.

\bibitem[{Dole} et~al.(2006)]{2006A&A...451..417D}
{Dole}, H., {Lagache}, G., {Puget}, J.-L., {Caputi}, K.~I.,
  {Fern{\'a}ndez-Conde}, N., {Le Floc'h}, E., {Papovich}, C.,
  {P{\'e}rez-Gonz{\'a}lez}, P.~G., {Rieke}, G.~H. \& {Blaylock}, M.
\newblock May 2006, \aap, 451, 417--429.

\bibitem[{Dye} et~al.(2006)]{2006ApJ...644..769D}
{Dye}, S., {Eales}, S.~A., {Ashby}, M.~L.~N., {Huang}, J.-S., {Webb}, T.~M.~A.,
  {Barmby}, P., {Lilly}, S., {Brodwin}, M., {McCracken}, H., {Egami}, E. \&
  {Fazio}, G.~G.
\newblock June 2006, \apj, 644, 769--777.

\bibitem[{Elbaz} et~al.(2002)]{2002A&A...384..848E}
{Elbaz}, D., {Cesarsky}, C.~J., {Chanial}, P., {Aussel}, H., {Franceschini},
  A., {Fadda}, D. \& {Chary}, R.~R.
\newblock March 2002, \aap, 384, 848--865.

\bibitem[{Elbaz}(2005)]{2005SSRv..119...93E}
{Elbaz}, D.
\newblock August 2005, Space Science Reviews, 119, 93--119.

\bibitem[{Farrah} et~al.(2006)]{2006ApJ...641L..17F}
{Farrah}, D., {Lonsdale}, C.~J., {Borys}, C., {Fang}, F., {Waddington}, I.,
  {Oliver}, S., {Rowan-Robinson}, M., {Babbedge}, T., {Shupe}, D., {Polletta},
  M., {Smith}, H.~E. \& {Surace}, J.
\newblock April 2006, \apjl, 641, L17--L20.

\bibitem[{Fixsen} et~al.(1998)]{1998ApJ...508..123F}
{Fixsen}, D.~J., {Dwek}, E., {Mather}, J.~C., {Bennett}, C.~L. \& {Shafer},
  R.~A.
\newblock November 1998, \apj, 508, 123--128.

\bibitem[{Gautier} et~al.(1992)]{1992AJ....103.1313G}
{Gautier}, III, T.~N., {Boulanger}, F., {Perault}, M. \& {Puget}, J.~L.
\newblock April 1992, \aj, 103, 1313--1324.

\bibitem[{Genzel} \& {Cesarsky}(2000)]{2000ARA&A..38..761G}
{Genzel}, R. \& {Cesarsky}, C.~J.
\newblock 2000, \araa, 38, 761--814.

\bibitem[{Giavalisco} et~al.(1998)]{1998ApJ...503..543G}
{Giavalisco}, M., {Steidel}, C.~C., {Adelberger}, K.~L., {Dickinson}, M.~E.,
  {Pettini}, M. \& {Kellogg}, M.
\newblock August 1998, \apj, 503, 543--+.

\bibitem[{Gonz{\'a}lez-Nuevo} et~al.(2005)]{2005ApJ...621....1G}
{Gonz{\'a}lez-Nuevo}, J., {Toffolatti}, L. \& {Arg{\"u}eso}, F.
\newblock March 2005, \apj, 621, 1--14.

\bibitem[{Grossan} \& {Smoot}(2007)]{2007A&A..4512G}
{Grossan}, B. \& {Smoot}, G.~F.
\newblock 2007, \aap, in press.

\bibitem[{Haiman} \& {Knox}(2000)]{2000ApJ...530..124H}
{Haiman}, Z. \& {Knox}, L.
\newblock February 2000, \apj, 530, 124--132.

\bibitem[{Hauser} et~al.(1998)]{1998ApJ...508...25H}
{Hauser}, M.~G., {Arendt}, R.~G., {Kelsall}, T., {Dwek}, E., {Odegard}, N.,
  {Weiland}, J.~L., {Freudenreich}, H.~T., {Reach}, W.~T., {Silverberg}, R.~F.,
  {Moseley}, S.~H., {Pei}, Y.~C., {Lubin}, P., {Mather}, J.~C., {Shafer},
  R.~A., {Smoot}, G.~F., {Weiss}, R., {Wilkinson}, D.~T. \& {Wright}, E.~L.
\newblock November 1998, \apj, 508, 25--43.

\bibitem[{Holland} et~al.(1998)]{1998astro.ph..9121H}
{Holland}, W.~S., {Cunningham}, C.~R., {Gear}, W.~K., {Jenness}, T., {Laidlaw},
  K., {Lightfoot}, J.~F. \& {Robson}, E.~I.
\newblock September 1998, ArXiv Astrophysics e-prints.

\bibitem[{Knox} et~al.(2001)]{2001ApJ...550....7K}
{Knox}, L., {Cooray}, A., {Eisenstein}, D. \& {Haiman}, Z.
\newblock March 2001, \apj, 550, 7--20.

\bibitem[{Knox}(1995)]{1995PhRvD..52.4307K}
{Knox}, L.
\newblock October 1995, \prd, 52, 4307--4318.

\bibitem[{Lagache} \& {Puget}(2000)]{2000A&A...355...17L}
{Lagache}, G. \& {Puget}, J.~L.
\newblock March 2000, \aap, 355, 17--22.

\bibitem[{Lagache} et~al.(2003)]{2003MNRAS.338..555L}
{Lagache}, G., {Dole}, H. \& {Puget}, J.-L.
\newblock January 2003, \mnras, 338, 555--571.

\bibitem[{Lagache} et~al.(2004)]{2004ApJS..154..112L}
{Lagache}, G., {Dole}, H., {Puget}, J.-L., {P{\' e}rez-Gonz{\' a}lez}, P.~G.,
  {Le Floc'h}, E., {Rieke}, G.~H., {Papovich}, C., {Egami}, E.,
  {Alonso-Herrero}, A., {Engelbracht}, C.~W., {Gordon}, K.~D., {Misselt}, K.~A.
  \& {Morrison}, J.~E.
\newblock September 2004, \apjs, 154, 112--117.

\bibitem[{Lagache} et~al.(2005)]{2005ARA&A..43..727L}
{Lagache}, G., {Puget}, J.-L. \& {Dole}, H.
\newblock September 2005, \araa, 43, 727--768.

\bibitem[{Lagache} et~al.(2007)]{2007ApJ...665L..89L}
{Lagache}, G., {Bavouzet}, N., {Fernandez-Conde}, N., {Ponthieu}, N., {Rodet},
  T., {Dole}, H., {Miville-Desch{\^e}nes}, M.-A. \& {Puget}, J.-L.
\newblock August 2007, \apjl, 665, L89--L92.

\bibitem[{Lahav} \& {Suto}(2004)]{2004LRR.....7....8L}
{Lahav}, O. \& {Suto}, Y.
\newblock July 2004, Living Reviews in Relativity, 7, 8--+.

\bibitem[{L{\'o}pez-Caniego} et~al.(2006)]{2006MNRAS.370.2047L}
{L{\'o}pez-Caniego}, M., {Herranz}, D., {Gonz{\'a}lez-Nuevo}, J., {Sanz},
  J.~L., {Barreiro}, R.~B., {Vielva}, P., {Arg{\"u}eso}, F. \& {Toffolatti}, L.
\newblock August 2006, \mnras, 370, 2047--2063.

\bibitem[{Magliocchetti} et~al.(2007)]{2007MNRAS.375.1121M}
{Magliocchetti}, M., {Silva}, L., {Lapi}, A., {de Zotti}, G., {Granato}, G.~L.,
  {Fadda}, D. \& {Danese}, L.
\newblock March 2007, \mnras, 375, 1121--1132.

\bibitem[{Marinoni} et~al.(2006)]{2006astro.ph.12123M}
{Marinoni}, C., {Le Fevre}, O., {Meneux}, B. \& {the VVDS team}.
\newblock December 2006, ArXiv Astrophysics e-prints.

\bibitem[{Matsuhara} et~al.(2000)]{2000A&A...361..407M}
{Matsuhara}, H., {Kawara}, K., {Sato}, Y., {Taniguchi}, Y., {Okuda}, H.,
  {Matsumoto}, T., {Sofue}, Y., {Wakamatsu}, K., {Cowie}, L.~L., {Joseph},
  R.~D. \& {Sanders}, D.~B.
\newblock September 2000, \aap, 361, 407--414.

\bibitem[{Miville-Desch{\^e}nes} et~al.(2001)]{2001phso.conf..471M}
{Miville-Desch{\^e}nes}, M.-A., {Abergel}, A. \& {Boulanger}, F.
\newblock {Dust Evolution in the Cold and Diffuse Interstellar Medium: The
  Herschel Perspective}.
\newblock In {Pilbratt}, G.~L., {Cernicharo}, J., {Heras}, A.~M., {Prusti}, T.
  \& {Harris}, R, editors, {\em ESA SP-460: The Promise of the Herschel Space
  Observatory}, pages 471--+, July 2001.

\bibitem[{Miville-Deschenes} et~al.(2007)]{2007arXiv0704.2175M}
{Miville-Deschenes}, M. ., {Lagache}, G., {Boulanger}, F. \& {Puget}, J. .
\newblock April 2007, ArXiv e-prints, 704.

\bibitem[{Negrello} et~al.(2007)]{2007astro.ph..3210N}
{Negrello}, M., {Perrotta}, F., {Gonzalez-Nuevo Gonzalez}, J., {Silva}, L., {De
  Zotti}, G., {Granato}, G.~L., {Baccigalupi}, C. \& {Danese}, L.
\newblock March 2007, ArXiv Astrophysics e-prints.

\bibitem[{Papovich} et~al.(2004)]{2004ApJS..154...70P}
{Papovich}, C., {Dole}, H., {Egami}, E., {Le Floc'h}, E.,
  {P{\'e}rez-Gonz{\'a}lez}, P.~G., {Alonso-Herrero}, A., {Bai}, L., {Beichman},
  C.~A., {Blaylock}, M., {Engelbracht}, C.~W., {Gordon}, K.~D., {Hines}, D.~C.,
  {Misselt}, K.~A., {Morrison}, J.~E., {Mould}, J., {Muzerolle}, J.,
  {Neugebauer}, G., {Richards}, P.~L., {Rieke}, G.~H., {Rieke}, M.~J., {Rigby},
  J.~R., {Su}, K.~Y.~L. \& {Young}, E.~T.
\newblock September 2004, \apjs, 154, 70--74.

\bibitem[{Puget} et~al.(1996)]{1996A&A...308L...5P}
{Puget}, J.-L., {Abergel}, A., {Bernard}, J.-P., {Boulanger}, F., {Burton},
  W.~B., {Desert}, F.-X. \& {Hartmann}, D.
\newblock April 1996, \aap, 308, L5+.

\bibitem[{Saunders} et~al.(1992)]{1992MNRAS.258..134S}
{Saunders}, W., {Rowan-Robinson}, M. \& {Lawrence}, A.
\newblock September 1992, \mnras, 258, 134--146.

\bibitem[{Steidel} et~al.(1998)]{1998astro.ph.12167S}
{Steidel}, C.~C., {Adelberger}, K.~L., {Dickinson}, M., {Giavalisco}, M. \&
  {Pettini}, M.
\newblock December 1998, ArXiv Astrophysics e-prints.

\bibitem[{Wang} et~al.(2006)]{2006ApJ...647...74W}
{Wang}, W.-H., {Cowie}, L.~L. \& {Barger}, A.~J.
\newblock August 2006, \apj, 647, 74--85.

\end{thebibliography}
\end{document}